\documentclass[review]{elsarticle}

\usepackage{lineno,hyperref}

\usepackage{graphicx}
\usepackage{mathtools}
\usepackage[dvips]{color}
\usepackage{epsfig}
\usepackage{amssymb}
\usepackage{amsmath}
\usepackage{bm}
\usepackage{float}
\usepackage{setspace}
\usepackage[margin=3cm]{geometry}

\journal{Journal of Theoretical Population Biology}

\biboptions{authoryear}

\begin{document}

\begin{frontmatter}

\title{Survival Probabilities at Spherical Frontiers}

\author[upenn]{Maxim O. Lavrentovich }

\address[upenn]{Department of Physics and Astronomy, University of Pennsylvania, Philadelphia PA 19104 USA}

\author[harvard]{David R. Nelson}

\address[harvard]{Lyman Laboratory of Physics, Harvard University, Cambridge MA 02138 USA}


\cortext[cor]{\textit{Email address:} {lavrentm@gmail.com} }

\begin{abstract}

Motivated by tumor growth and spatial population genetics, we study the interplay between evolutionary and spatial dynamics at the surfaces of three-dimensional, spherical range expansions.   We consider range expansion radii that grow with an arbitrary power-law in time:  $R(t)=R_0(1+t/t^*)^{\Theta}$, where $\Theta$ is a growth exponent, $R_0$ is the initial radius, and $t^*$ is a characteristic time for the growth, to be affected by the inflating geometry.  We  vary the parameters $t^*$ and $\Theta$ to capture a variety of possible growth regimes.   Guided by recent results for
two-dimensional inflating range expansions, we identify key dimensionless parameters that describe the survival probability of a mutant cell with a small selective advantage arising at the population frontier.  Using analytical techniques, we calculate this probability for arbitrary $\Theta$.  We compare our results to  simulations of linearly inflating expansions ($\Theta=1$ spherical  Fisher-Kolmogorov-Petrovsky-Piscunov waves) and treadmilling populations ($\Theta=0$, with cells in the interior removed by apoptosis or a similar process).  We find that mutations at linearly inflating fronts have  survival probabilities enhanced by factors of 100 or more relative to mutations at treadmilling population frontiers.  We also discuss the special properties of ``marginally inflating'' $(\Theta=1/2)$ expansions.

\end{abstract}

\begin{keyword}
survival probability; genetic drift; range expansions; avascular tumor evolution; selection\end{keyword}

\end{frontmatter}


\section{\label{SIntro}Introduction}

 Early tumor evolution is driven by rare driver mutations that sweep the prevascular tumor population at the frontier and push the growing cell mass further down the path toward metastasis.  Hence, an understanding of the evolutionary dynamics governing the survival of such mutations is crucial in cancer prevention \citep{cancerEvo,cancerEvo2}.  One significant, largely unexplored aspect of this evolution is the effect of tumor geometry.
 An important \textit{in vitro} model of cancer is the multicellular tumor with an approximately spherical shape, or ``spheroid''.  The spheroid captures many of the essential features of solid tumors \textit{in vivo} and is a model for anti-cancer therapies \citep{spheroids1, spheroids2, spheroids3}.  Spheroids are especially useful for understanding small, avascular tumors.  In the later stages of growth, in order for the tumor to survive, it requires a vascular system and undergoes angiogenesis \citep{angiogenesis}.  The growth then becomes more complicated, and  more sophisticated modeling efforts are necessary \citep{angiomodel2,angiomodel1}. We focus here on  the earlier evolutionary dynamics of spheroidal range expansions in two and three dimensions.   We assume that attractive cell-cell interactions keep such aggregates approximately spherical, i.e. that there is an effective surface tension, similar to that observed for yeast cell colonies \citep{brenneryeast}.  Although we are motivated by tumor evolution, our models are intended to be quite general.  Two-dimensional and three-dimensional expansions  may be realized in experiments, for example, using microbial or yeast populations in hard and soft agar, respectively \citep{KorolevMueller, KorolevBac, MKNPRE}.

We will be particularly interested in computing the survival probability of a mutation that occurs among the dividing cells at the surface of a spherical or circular population of initial radius $R(t=0)=R_0$, which may or may not increase in time.  In general, the radius $R(t)$ has a complicated time dependence, especially in tumor growth.  At the early stages, cells divide everywhere inside the tumor, and the cluster radius grows exponentially in time.   After the tumor reaches a size larger than a nutrient shielding length  \citep{MLDRNNut}, nutrients will no longer be able to diffuse into the tumor interior.  This effect, combined with inward pressure from the surrounding non-cancerous tissue \citep{treadmill1, treadmill2, treadmill3}, decreases the growth rate toward the center of the tumor.  The radius $R(t)$ then grows more slowly.  We will model the growth generally  as
\begin{equation}
R(t) = R_0 \left[1 + \frac{t}{t^*} \right]^{\Theta}, \label{eq:radiusgrowth}
\end{equation}
 where $R_0$ is the initial tumor radius, $\Theta$ is a (possibly time-dependent) growth exponent, and $t^*$ is a characteristic time for the power-law growth in an inflating geometry.  Both $\Theta$ and $t^*$ may be tuned to model various growth regimes.  For example, for a substantial portion of the growth in tumors, the radius grows linearly in time $(\Theta =1)$, and  $t^* = R_0/v$, where  $v$ is the front speed \citep{tumorg}.  Linear growth and nutrient shielding are also present in microbial populations grown in Petri dishes \citep{KorolevMueller}.  Equation~(\ref{eq:radiusgrowth}) can also model an exponential growth regime $R(t)=R_0 e^{\lambda t}$ with rate $\lambda$ if we let both $\Theta,t^* \rightarrow \infty$, such that $\lambda = \Theta/t^*$ is held constant.

 \begin{figure}[H]
\centering
\includegraphics[width=2.5in]{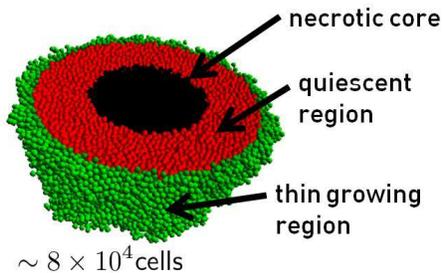}
\caption{\label{fig:TumorScheme}  (Color online)  A schematic of a treadmilling tumor.  Due to nutrient shielding, cells divide in a thin green region at the frontier.  In the red region, cells are in an arrested state and do not grow.  In the necrotic core, cells undergo apoptosis and their contents are flushed out of the cluster, resulting in an overall volume loss.  This volume loss can balance the gain of volume at the cluster periphery, resulting in a ``treadmilling'' effect and a cell mass with a constant radius \citep{treadmill1,treadmill2,treadmill3, pottscancer}.  }
\end{figure} 

 Eventually, apoptosis may be induced at the tumor center, creating a necrotic region \citep{treadmill1,treadmill2,treadmill3}, illustrated in Fig.~\ref{fig:TumorScheme}.  The cells at the tumor periphery continue to divide relatively rapidly.  Thus, a ``treadmilling'' effect is created, and the tumor experiences a rapid turnover of cells at its surface while remaining the same size, a situation we represent by a growth exponent $\Theta =0$ in Eq.~(\ref{eq:radiusgrowth}). We will show that the different growth regimes captured by varying $\Theta$ have dramatically different consequences for the fate of mutations at the tumor frontier.   We will focus on $\Theta=0$, $\Theta=1$, and $\Theta=1/2$, capturing, respectively, treadmilling, linear inflation, and an intriguing borderline growth regime.

\begin{figure}[H]
\centering
\includegraphics[height=2.3in]{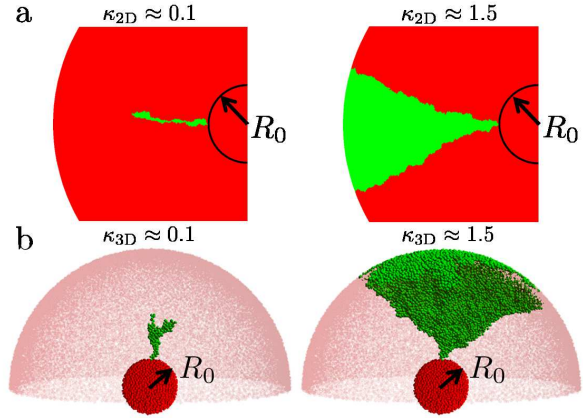}
\caption{Examples of simulated mutant clusters (green cells) in two- and three-dimensional range expansions (see Section~\ref{SModel}), generated using two different values of the key dimensionless selection parameters $\kappa_{\mathrm{2D},\mathrm{3D}}$ defined in Eq.~(\ref{eq:kappasummary}). (a) Circular range expansions with a uniform front with an initial radius $R_0=50$ average cell diameters and a single initial mutant green cell at the population frontier. (b) Spherical range expansions with uniform fronts and a single green cell at the initial population frontier with radius $R_0 = 10$ cell diameters. \label{fig:kappapics}}
\end{figure}

  The actively growing region in a tumor mass or a spherical microbial population  can be quite thin, with a width of just a few cell diameters  \citep{hochberg, MLDRNNut}.  In this case, genetic drift is strong and  can locally fix the mutation at the population frontier.  This local fixation creates a mutant ``sector,'' i.e., a region along the front that is entirely occupied by the mutant cells.  Example sectors, marked in green, are shown in Fig.~\ref{fig:kappapics} for circular and spherical range expansions.   Previous studies have focused on the deterministic movement of these mutant sectors: The sectors inflate or deflate due to a mutant selective advantage or disadvantage, respectively  \citep{cancernowak}. However, genetic drift will introduce fluctuations in the sector motion at its boundaries that can drive the mutation to extinction, as illustrated on the left panels of Fig.~\ref{fig:kappapics}(a) and Fig.~\ref{fig:kappapics}(b).  For example, in circular expansions [Fig.~\ref{fig:kappapics}(a)], the sector has two boundaries which both perform random walks, as observed in microbial range expansions \citep{KorolevRMP, KorolevMueller}.   Selection introduces a bias to the sector boundary motion.  Also, the increasing population radius $R(t)$ will deterministically increase the distance between the boundaries.      If the sector boundaries collide, the sector vanishes, and the mutation goes extinct [left panel of Fig.~\ref{fig:kappapics}(a)].    

Because the two  boundaries of mutant sectors in two-dimensional expansions perform random walks, the distance between the boundaries performs a random walk as well.  The mutant survival probability, then, is the probability that this distance never vanishes, i.e., that the random walk it performs never reaches the origin.   Such a probability is a well-studied first-passage property of a random walk \citep{redner}.   Previous studies, such as \citet{nelsonhallatschek, KorolevRMP, MKNPRE}, have exploited these known random walk results to calculate survival probabilities of mutations in two-dimensional populations.   We will review some of these previous results for two-dimensional expansions and generalize them to arbitrary growth exponents $\Theta$.  We will then use them to motivate our discussion of three-dimensional expansions. 

 In three-dimensional populations, the focus of the present work, the mutant sector can have a complex, branched shape, as shown in Fig.~\ref{fig:kappapics}(b). Characterizing the boundary positions with just two random walks is impossible in this case.  Although we can still treat mutant cell lineages using random walks  [see, e.g., \cite{clusteringvoter} or the chapter on voter models in \citet{voter1}],  incorporating selection is more complicated than in the two-dimensional case \citep{bramson}. Also, we know of no generalization to inflating frontiers.   So, instead of mapping to random walks,   we study the time evolution of the coarse-grained density of mutant cells along the population frontier of spherical range expansions.  We then apply a field-theoretic analysis of the time evolution that allows us to treat genetic drift, selection, and inflation within a single theoretical framework.

 By using random walk theory for two-dimensional range expansions and field-theoretic techniques for three-dimensional ones,  we will show that the key dimensionless parameters for mutant survival  are, respectively,
\begin{equation}
 \kappa_{\mathrm{2D}}  =s \sqrt{\dfrac{  t^*}{ \tau_g}} \quad \mbox{ and } \quad
 \kappa_{\mathrm{3D}}  = \dfrac{st^*}{\tau_g},
 \label{eq:kappasummary}
\end{equation} 
where $t^*$ is the characteristic time of the radius growth defined in Eq.~(\ref{eq:radiusgrowth}), $s$ is the selective advantage, and $\tau_g$ is a generation time.  The mutant survival will also depend on the initial number of mutant cells and the growth exponent $\Theta$.  The mutant cluster shapes at different $\kappa_{\mathrm{2D},\mathrm{3D}}$ for linearly inflating frontiers [$\Theta=1$ in Eq.~(\ref{eq:radiusgrowth})] are  illustrated in Figs.~\ref{fig:kappapics}(a) and \ref{fig:kappapics}(b) for two- and three-dimensional expansions, respectively.   For $\kappa_{\mathrm{2D},\mathrm{3D}} \gtrsim 1$,\ the selective advantage is large enough to sustain  a surviving mutant cluster after a short time, allowing it to survive indefinitely and (if no other mutant sectors are present) to  sweep the population deterministically.  When $\kappa_{\mathrm{2D},\mathrm{3D}} \ll 1$, however, the effects of genetic drift are important    for a longer time, and can lead to an extinction of the cluster.  

The remainder of the paper is organized as follows:  We introduce the simulation model used to create two- and three-dimensional expansions in Section~\ref{SModel}. We impose a global constraint that insures compact clusters, as a computationally efficient way of emulating the effect of surface tension.  In Section~\ref{SPopGen} we review the deterministic dynamics of mutant sectors, exemplified by the average sector shapes corresponding to the right panels of Figs.~\ref{fig:kappapics}(a) and \ref{fig:kappapics}(b). We also introduce the stochastic equation governing both the deterministic and stochastic dynamics. In Section~\ref{S2D} we review and extend results for two-dimensional expansions, as these provide insights into the three-dimensional case relevant to tumors.  In Section~\ref{S3D} we calculate the survival probability of mutations in three dimensions, relegating details to \ref{appx:responsefunc} and \ref{appx:survp}.    Results for the case of $s<0$ with linear inflation are presented in \ref{appx:negsel}.  The smooth circular and spherical fronts generated in our simulations facilitate comparison with analytical results.  We make concluding remarks in Section~\ref{SConclusions}.

\section{\label{SModel} Simulations}

   Our simulations  will focus on populations with a single layer of growing cells.    We also assume, to simplify the analysis, \textit{compact} fronts where the population front closely approximates a uniform circle or sphere at all times.  The latter approximation will be valid as long as front undulations and the selective advantage $s$ are small. (For discussions of how the dynamics change when these approximations are not valid, see \citet{cancernowak,KorolevMueller,frey}).    As discussed in more detail by \citet{MKNPRE}, these approximations have the advantage of allowing us to perform a ``dimensional-reduction'' by focussing on just the dynamics at the population frontier, which will be effectively one- and two-dimensional inflating geometries for two- and three-dimensional expansions, respectively. 

All along these frontiers, cells will compete locally to divide into new territory and form the next generation of cells.  We consider two types of cells: a wild-type red cell and a green cell with a selectively advantageous ``driver'' mutation. The mutant, ``driver'' green cells have a base growth rate $\Gamma_g $ that is greater than the red wild-type cell growth rate $\Gamma_r$ by a factor of $\Gamma_g/\Gamma_r=1/(1-s)$, where $s$ is a selective advantage.    The difference in the growth rates $\Gamma_g/\Gamma_r$ determines the probability $p_g$ of a green offspring cell occupying an available spot along the frontier.  Specifically, if there are a total of $z$ ``parent'' cells, of which $n_g$ are green, competing to divide into an available spot, the probability $p_g=p_g(n_g,z)$  of placing a green offspring in the spot  is 
\begin{equation}
p_g(n_g,z) =\frac{(\Gamma_g /\Gamma_r)n_g}{z-n_g+(\Gamma_g /\Gamma_r)n_g}= \frac{n_g}{(1-s)z+sn_g}. \label{eq:updateprobs}
\end{equation}
New cells are placed one at a time in predefined locations to create expansions with uniform fronts, as described below.  The probability of a particular cell color dividing into the new spot is determined by Eq.~(\ref{eq:updateprobs}), with the probability of a red offspring given by $p_r(n_g,z)=1-p_g(n_g,z)$.

\begin{figure}[H]
\centering
\includegraphics[width=2.5in]{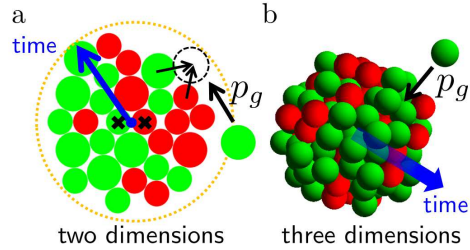}
\caption{\label{fig:smallclusters}  Schematics of simulated radial range expansions in (a) two and (b) three dimensions.  In (a), cells of two different sizes compete to divide into new territory. The possible cell locations are chosen in advance using an algorithm described in the main text and in \citet{MKNPRE}.   The algorithm is seeded with the two cell locations indicated by black crosses.  The two  cell sizes and their relative abundances are chosen to create an amorphous, isotropic cell packing.   In (b), cells with identical diameters $a$ compete to divide.  The cell locations are generated   using the Bennett model \citep{Bennett}.  In both (a) and (b), the virgin territory that is closest to the cell cluster center is settled first to create an effect similar to a surface tension. In both models, the green cells out-compete the red with probability $p_g$ given by Eq.~(\ref{eq:updateprobs}). }
\end{figure}

We fill empty sites  one at a time, starting with the site that is closest to a central reference point, as shown in Fig.~\ref{fig:smallclusters}.  In one time step, the cells adjacent to the empty site compete  to divide into the site.      Repeating this process creates a growing cell cluster with a relatively smooth circular or spherical front, approximating the effect of a surface tension. Surface tension will suppress both front undulations and prominent bulges arising due to the faster growth of the mutant cells. Our methodology is reasonable provided surface tension forces only move daughter cells a short distance after they are born (i.e. about a cell diameter).       The resulting compact cell cluster could also arise due to biological reasons. For example,  cells may secrete a chemical that promotes cell divisions, which then are only feasible at the cluster surface where there is room to divide \citep{MKNPRE}. This smooth frontier approximation is not only more tractable analytically, but increases our simulations' computational efficiency.

 The set of empty sites into which cells may divide (i.e. the set of possible cell positions) is generated in advance.  One possibility is to arrange the cell positions in a lattice.  However, for outwardly growing circular range expansions, this approach produces strong artifacts of the lattice symmetry  in the evolutionary dynamics \citep{MKNPRE}.  We mitigate these effects by generating an amorphous packing of empty sites using a modified Bennett model \citep{rubinstein}.  In this model, we first seed the packing with two adjacent  sites, indicated by black crosses in Fig.~\ref{fig:smallclusters}(a).   The center of this initial seed is our central reference point. Then,  cell locations (sites) are generated one at a time and are placed as close as possible to the central reference point (without overlapping previously generated sites). Each site must also touch at least two previously placed sites. A smaller site is chosen over the larger with some specified probability.  \citet{MKNPRE}  found that isotropic, amorphous cell packings occurred if cells are chosen with a small-to-large radius ratio of $\rho =8/11$, with the smaller one placed with 60\% probability, choices we also make here.   Finally, note that evolutionary dynamics do not depend on the cell size.  A small cell may divide into an empty site that fits a large cell, or vice versa.

 For spherical range expansions, we generate our set of possible cell locations  using the original Bennett model \citep{Bennett} with identical spherical cells with diameter $a$.    These cells are now placed closest to the central reference point such that each touches at least three previously placed cells.   The initial seed is a tetrahedron of cells.  The algorithm creates an amorphous packing without the need to vary the cell size as in the two-dimensional case.  Circular and spherical expansions with relatively small population sizes are shown in Fig.~\ref{fig:smallclusters}.  Note that  a cell diameter is more complicated to define in the two-dimensional expansion because we use both small and large cells.  So,   $a$  in this case will denote an effective  spacing between cells that is approximately equal to the average cell diameter.

To further mitigate lattice artifacts, we initialize the mutant green cells during each simulation run in one of at least 100 different, random locations,  sampled randomly and uniformly along the population periphery.   In two dimensions, all cells along the initial, circular population front have equal probability of being chosen as the mutant green cell. Thus, the mutant may be either a small cell or a large cell, depending on the location chosen.  In three dimensions, we first perform a random rotation of the initial spherical population as described by \citet{Arvo92fastrandom}. We then seed the mutated, green cells at the north pole of the spherical population.   This ensures a uniform, random sampling of the initial population front.  We study initial fronts with just one green cell and fronts with a localized, compact clumps of many green cells.

This algorithm generates linearly inflating fronts with $\Theta=1$.  It is possible to simulate \textit{treadmilling} tumors with $\Theta =0$ by adding a step that evolves the population backwards, toward the interior of the cluster:   First, we evolve the range expansion \textit{forward} just a few generations, generating a cluster of radius $R_0+ n v \tau_g \approx R_0+ na$, where $n\approx 2.5$.
Second, we go back to the cells at approximately a distance  $R_0+1.5a$  from the cluster center, and replace them with new offspring cells.    This is done by evolving the population \textit{backwards}, toward the central reference point.  The cells at the population periphery (at distances $R_0+na$ from the center) divide toward the population interior and replace the cells generated during the forward sweep.   The previously generated cells are replaced one at a time as the population evolves deeper into the interior.   This replacement models both the  volume loss due to  apoptosis  at the population center and pressure from the surrounding tissue, which prevents the tumor from growing outward.     This backwards evolution is continued until a population front with radius $R_0-nv \tau_g$ is created.    Finally, we go back to the cells at radius $R_0$ and repeat the first step,  re-evolving the initial population radius outward up to a radius $R_0+nv \tau_g$.  We  again  let cells divide into sites previously occupied by cells generated during the backward sweep.   The backwards and forward sweeps are repeated over and over to model a turn-over of cells at the tumor surface, taken to be the spherical shell of cells with fixed radius $R_0$ and thickness $v \tau_g \approx a$.

In both the treadmilling and expanding populations, our cells may divide as long as they are adjacent to either an empty site or, in the treadmilling case, a previously generated cell chosen to be replaced.  So, the lifetime of a cell is entirely governed by the availability of adjacent empty sites.  Once these have been filled, the cell may no longer divide.

\section{\label{SPopGen}  Population genetics at compact population fronts}

\begin{figure}[H]
\centering
\includegraphics[width=2.5in]{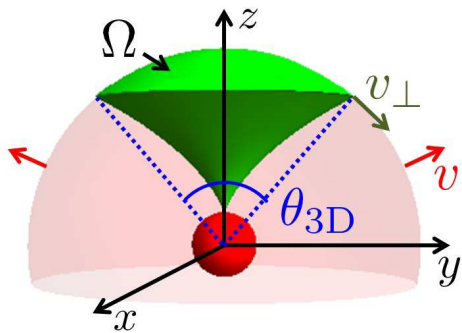}
\caption{\label{fig:3dlogspiral}  Schematic of the deterministic motion of a mutant sector arising at the surface of a spherical population of red cells.  The green sector invades the red population with a selection-dependent lateral velocity $v_{\perp} = v w_{\mathrm{3D}}(s)$, where $v$ is the radial front speed.  The sector forms a logarithmic cone, formed by rotating a logarithmic spiral curve parameterized by the angle $\theta_{\mathrm{3D}}$ around the $z$-axis.  In our simulations, we ignore possible bulges in the population front and  neglect the enhanced velocity of the green sector relative to the red domain expansion velocity $v$.  Hence, the green sector  forms a spherical cap with some solid angle $\Omega$ at the spherical population frontier. }
\end{figure}

  Before we analyze survival probabilities of mutations, it is instructive to first consider the deterministic dynamics of a mutation.   In particular, we will now describe how a mutation with a selective advantage  sweeps across the population frontier in a radial range expansion [corresponding to $\Theta=1$ in Eq.~(\ref{eq:radiusgrowth})], ignoring sector extinction due to genetic drift \citep{cancernowak,KorolevMueller}.  A surviving mutant forms a sector as shown in Fig.~\ref{fig:kappapics} for two dimensions and schematically in Fig.~\ref{fig:3dlogspiral} for three dimensions.     We assume $s \ll 1$ and/or a strong surface tension, so that we may ignore any bulges created by the green strain and to make contact with our simulations.    The boundary of this sector is a genetic Fisher wave in which the mutant green strain ``invades'' the red wild-type region azimuthally along the frontier with some speed $v_{\perp}$.  This means that in a two-dimensional circular expansion, the two sector boundaries will move away from each other with a relative speed $2 v_{\perp}$ along the population frontier.  The range expansion itself will grow outward with a velocity $v$.  Hence, the sector angle $\theta_{\mathrm{2D}}$ between the two boundaries (relative to the center of the range expansion) will increase as $ d \theta_{\mathrm{2D}}/dt=2 v_{\perp}/R(t)$, where $R(t) = R_0+vt$.  Integrating this differential equation, we find that the angular sector size $\theta_{\mathrm{2D}}(t)$ is given by
\begin{equation}
 \theta_{\mathrm{2D}}(t) = w_{\mathrm{2D}}(s) \ln \left[ 1+\frac{vt}{R_0} \right]+\theta_0, \label{eq:2dlogspirals}
\end{equation}
where $\theta_0$ is the initial sector size and $w_{\mathrm{2D}}(s)=2v_{\perp}/v$ is a rescaled sector boundary speed (Fisher wave speed) that depends on the selection coefficient $s$ and is influenced by the strength of the genetic drift.   Equation~(\ref{eq:2dlogspirals})  describes a sector with boundaries that form logarithmic spirals that spiral away from each other  \citep{KorolevMueller}.

  In general,  $w_{\mathrm{2D}}(s) \rightarrow 0$ as $s \rightarrow 0$ in Eq.~\ref{eq:2dlogspirals}, independently of the strength of the genetic drift.  For our simulations,  $w_{\mathrm{2D}}(s) \approx 1.2a s/(v \tau_g) $, where $v \tau_g\approx a$ is the distance the front moves in one generation and $a$ is the effective spacing between cells discussed in  Sec.~\ref{SModel}.  The scaling    $w_{\mathrm{2D}}(s) \sim s$  is a consequence of strong genetic drift, and, as explored in more detail by \citet{MKNPRE}, can be derived by mapping the sector boundaries to random walks (see Sec.~\ref{S2D}).   The mapping of sector boundaries to random walks allows us to generalize our analysis to sectors with bulges, as well.  Indeed, we expect the green sector to bulge out because it will grow out radially faster than the red strain.  However, provided the front has some effective line tension and does not roughen, the sector boundaries will still perform biased random walks with some effective rescaled bias speed  $w_{\mathrm{2D}}(s) \sim s$ in the presence of bulges.  Hence, our analysis of the sector survival probability in Section~\ref{S2D} will apply for this case, as well.  We will now turn to sector shapes in three dimensions.  In three dimensions, mutations with a strong selective advantage form  ``logarithmic cone'' sectors, i.e. solids of revolution formed by logarithmic spirals \citep{cancernowak}.

\begin{figure}[H]
\centering
\includegraphics[height=2.6in]{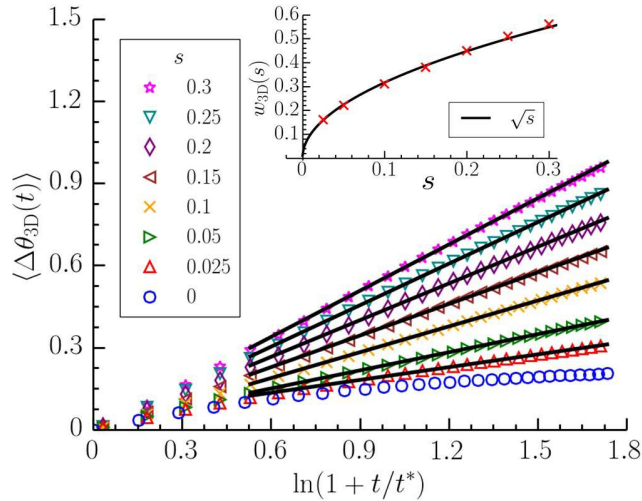}
\caption{ Simulated mutant clusters in a spherical range expansion ($R_0=30$ cell widths) with an initial solid angle $\Omega_0\approx 0.20$ steradians.  We plot the angle $\theta_{\mathrm{3D}}(t)$ delineating the edge of the conical shape formed by the mutant cluster edge, averaged over 38400 surviving sectors.  After an initial transient,  mutant sectors form logarithmic cones for any selective advantage $s>0$, as illustrated by the linear fits through the data (black lines).  In the inset, we plot the slope of these lines, $w_{\mathrm{3D}}(s)$.  (Note that the case $s=0$ at the bottom is exceptional.) We find, to an excellent approximation, $w_{\mathrm{3D}}(s) \approx \sqrt{s}$.  \label{fig:logspirals}}
\end{figure}

       We can compute the analogous angle $\theta_{\mathrm{3D}}$ for a spherical expansion by calculating the solid angle $\Omega=\Omega(t)$ covered by the mutant sector at time $t$, as shown in Fig.~\ref{fig:3dlogspiral} \citep{cancernowak}.  Suppose that the initial solid angle covered by the mutant sector is $\Omega_0$. The corresponding initial angle is $\theta_0 = 2 \arccos(1-\Omega_0/2\pi)$.  Then, the angle of the shape of the edge of the logarithmic cone is given by the relative angle $\Delta \theta_{\mathrm{3D}}(t)=\theta_{\mathrm{3D}}-\theta_0$: 
\begin{equation}
\Delta \theta_{\mathrm{3D}}(t) =2\left[ \arccos\left(1-\frac{\Omega}{2\pi}\right)-\arccos\left(1-\frac{\Omega_0}{2\pi}\right) \right]= w_{\mathrm{3D}}(s) \ln \left[ 1+\frac{t}{t^*} \right], \label{eq:det3dspiral}
\end{equation} where  $w_{\mathrm{3D}}(s) = 2v_{\perp}(s)/v$. We check this result with simulations in Fig.~\ref{fig:logspirals}. We find that Eq.~(\ref{eq:det3dspiral}) accurately describes the average mutant sector shape, as shown by the linear fits through the data in the main plot.  Note that when $s=0$, the logarithmic cone sector boundary collapses, and the sector boundary  approaches a constant solid angle.  When many mutations with $s>0$ are present, these logarithmic cones of mutations can collide to form interesting limiting shapes discussed in detail by \citet{cancernowak} (see also \citet{MartensHallatschek}).  Upon using Eq.~(\ref{eq:det3dspiral})  to extract the rescaled lateral Fisher wave velocity $w_{\mathrm{3D}}(s)$,  we find $w_{\mathrm{3D}}(s) \approx \sqrt{s}$ to an excellent approximation, as illustrated in the inset of Fig.~\ref{fig:logspirals}.  Note that this square-root scaling is markedly different from the circular expansion case.  Three-dimensional inflating range expansions are thus consistent with the classic  result that the Fisher wave-speed should approach $v_{\perp}= 2 \sqrt{D s}$ when there is no genetic drift, where $D$ is a spatial diffusion coefficient (see \citet{fisher} and discussion below).  We expect that the coefficient of the $\sqrt{s}$ term  depends on the noise (genetic drift) strength, which is fixed in our simulations. Note that $v_{\perp}(s)$ for small $s$ is much bigger than the difference between the mutant and wild-type radial expansion velocities (of order $s$), thus justifying our neglect of the bulge in three dimensions. This scaling result is consistent with previous studies of noisy Fisher waves in higher dimensions \citep{highdnoisyfisher, strongnoisefisher}. 

  We now have to introduce genetic drift explicitly to take into account possible sector extinction.  This will be done in Sec.~\ref{S2D} for two-dimensional expansions by mapping sector boundaries to random walks.  However,  to better understand our techniques for three-dimensional expansions in Sec.~\ref{S3D},   we  now develop some additional  results for noisy Fisher waves in non-inflating range expansions.  We define a coarse-grained green cell (mutant) fraction $f \equiv f(\mathbf{x},t)$ at positions $\mathbf{x}$ along the front at time $t$.  We compute this fraction over box-shaped regions of volume $\ell^d$, where $\ell$ is the box length or effective lattice spacing, and $d$ is the spatial dimension of the front, i.e., $d=1$ for a two-dimensional expansion and $d=2$ for a three-dimensional expansion.  To compare the analytic results with  our simulations where the fronts are monolayers of closely-packed cells, $\ell$ is chosen such that $\ell^d$ describes a region of the front occupied by  one cell.  For example, for a one-dimensional frontier in which the cells are collinear and touching each other, $\ell = a$ is a cell diameter. We assume that the characteristic size of a mutant cluster is small compared to the population front radius $R_0$ and that the front is uniform and flat. If the front is also not inflating ($\Theta=0$),  the coarse-grained green cell fraction $f \equiv f(\mathbf{x},t)$  (upon exploiting dimensional reduction) obeys the Langevin  equation
\begin{equation}
\partial_t f = D \nabla_{\mathbf{x}}^2 f+s \tau_g^{-1} f(1-f)+\sqrt{2\Delta \ell^d   \tau_g^{-1} f\left(1-f \right)} \, \xi(\mathbf{x},t), \label{eq:linlang}
\end{equation}
where $s$ is the selective advantage of the mutant, $D$ is a spatial diffusivity, $\tau_g$ is the generation time, $\Delta$ is a genetic drift strength, and $\xi$ is a Gaussian, white noise [$\langle \xi(\mathbf{x},t) \xi(\mathbf{x}',t') \rangle=\delta(\mathbf{x}-\mathbf{x}') \delta(t-t')$] interpreted in the It\^o sense (see Section 4.3 in \citet{gardiner} or \citet{KorolevRMP}).  The strength $\Delta$ of the genetic drift scales like $ N^{-1}$, where the effective population size $N$ is approximately the number of organisms in each region $\ell^d$ that compete to divide at the population frontier.  Equation~(\ref{eq:linlang}) may also be derived using a stepping stone model \citep{KorolevRMP}.

 If we remove the genetic drift term in Eq.~(\ref{eq:linlang}) by setting $\Delta=0$, we recover the Fisher-Kolmogorov-Petrovsky-Piscunov equation \citep{fisher, KorolevRMP}.  This equation allows traveling-wave solutions that describe the deterministic sweep of a mutant sector \citep{KorolevMueller}. These Fisher waves move at speed $v = 2 \sqrt{Ds}$.    The genetic drift induced by a non-zero $\Delta > 0$ can  strongly influence these traveling-wave dynamics \citep{strongnoisefisher,doeringduality} and can extinguish an embryonic mutant sector.     There are mathematical difficulties associated with interpreting Eq.~(\ref{eq:linlang})  for $d=2$ when the noise term is present $(\Delta >0),$ and when a continuous variable $\mathbf{x}$ is used \citep{bartonstepping}.  However, we will always impose finite large and small distance cutoffs ($R_0$ and $a$, respectively) where necessary to get meaningful results, and we will test and confirm our analytic results with careful numerical simulations.  So, although the sense in which Eq.~(\ref{eq:linlang})  is mathematically sensible  and may be used to compute averages and probabilities for arbitrary $d$ has been explored in the physics literature \citep{timthesis,CDPFT}, understanding this fascinating, subtle problem is not relevant for the calculations and regimes explored in our paper.   Also, to understand the evolutionary dynamics of inflating circular and spherical range expansions in the presence of genetic drift ($\Delta>0$) using Eq.~(\ref{eq:linlang}), the equation has to be appropriately modified to take into account the inflating population fronts.  We will make the appropriate modifications in  Sec.~\ref{S3D}, where we will also analyze the equation with field-theoretic techniques.

 We  will focus on an important, biologically relevant consequence of Eq.~(\ref{eq:linlang}): What is the probability that a single green mutant cell establishes a surviving cluster at long times, given an initial  green cell fraction $f(\mathbf{x},t=0)$ at the population frontier? In the strong genetic drift limit, \citet{doeringduality} mapped Eq.~(\ref{eq:linlang}) in one dimension to a reaction-diffusion model  and found that  the survival probability for the mutants is
\begin{equation}
P[s \geq 0,\Delta; f(x,t=0)]= 1- \exp \left[- \frac{s}{\Delta \ell} \int \mathrm{d}x\, f(x,t=0) \right], \label{eq:linsurvp}
\end{equation} 
 where we integrate $f(x,t=0)$ over all positions $x$ along the population frontier.    Equation~(\ref{eq:linsurvp}) is valid for non-inflating, large population frontiers. We expect Eq.~(\ref{eq:linsurvp}) to hold for three-dimensional expansions, as well.   This hypothesis was tested for a more sophisticated range expansion model developed by \citet{pigolotti}. We will show that Eq.~(\ref{eq:linsurvp}) can be derived via a field theoretic technique in Section~\ref{S3D}. Note that our choice of effective lattice spacing $\ell$ will allow us to relate the integral over the initial fraction $f(\mathbf{x},t=0)$ to the initial number of mutant green cells $N_0$:
\begin{equation}
N_0 =  \frac{1}{\ell^d} \int \mathrm{d}^d \mathbf{x} \, f(\mathbf{x},t=0), \label{eq:initialnummut}
\end{equation}
 where we may consider both one-dimensional $(d=1)$ and two-dimensional $(d=2)$ population fronts.  Substituting Eq.~(\ref{eq:initialnummut}) into Eq.~(\ref{eq:linsurvp}) for $d=1$, we see that   the survival probability vanishes linearly with $s$, $P \sim s\Delta^{-1} N_0$, when  $s \ll (\Delta \ell)^{-1} N_0$.   In contrast, we will find in Sections~\ref{S2D} and \ref{S3D} that for inflating expansions in two and three dimensions, this probability instead approaches a non-zero value as $s \rightarrow 0$.

The $s<0$ case, although not the focus of this paper, is  interesting and subtle.  In the deterministic regime, the boundaries of the sector in two dimensions form logarithmic spirals  that spiral toward each other for $s<0$, forming a tongue-like, or flower petal shape.  Similarly, in three dimensions, the boundary of the logarithmic cone spirals inward instead of outward.  Hence, we expect that the boundaries of any surviving mutant cluster will eventually be pinched off due to this deterministic motion for $s<0$.  Our theory predicts that the only way a mutant sector with $s<0$ can survive at long times is if it wraps all the way around the circumference of the population.  Otherwise, the long time survival probability vanishes: $P_{\infty}(s<0)=0$.  However, the eventual extinction of a sector might take a very long time, due to the logarithmically slow deterministic dynamics.  The convergence of the survival probability to its long-time steady-state value is extremely slow for $s<0$. Thus, we may not always be able to probe this long-time limit in our simulations, which can only evolve the range expansions up to some finite time. We briefly examine some of these issues in \ref{appx:negsel}.

\section{\label{S2D} Treadmilling and Expanding Populations in Two Dimensions}

In two-dimensional, circular populations, mutant sector boundaries perform random walks along the population frontier \citep{KorolevRMP, KorolevMueller}.  Each mutant sector will have two boundaries, as shown  in Fig.~\ref{fig:kappapics}.  Since each sector boundary performs a random walk, the sector angle between them,  $\theta_{\mathrm{2D}}(t)$, performs a random walk, as well.  We can write down a general stochastic differential equation  for $\theta_{\mathrm{2D}}$ that describes its random walk on a circular frontier    with a time-dependent radius $R(t)$ given by Eq.~(\ref{eq:radiusgrowth}):
\begin{equation}
\frac{d}{d t}\,\theta_{\mathrm{2D}}(t)  =  \frac{ V_{\mathrm{2D}}}{  (1+t/t^*)^{\Theta}}  + \sqrt{\frac{2D_{\mathrm{2D}}}{(1+t/t^*)^{2 \Theta}}} \, \eta(t), \label{eq:2dangleSDE}
\end{equation}
where $V_{\mathrm{2D}}$ is a selection-dependent bias and  $D_{\mathrm{2D}} $ is a  diffusion constant.      The  function $\eta(t)$ in Eq.~(\ref{eq:2dangleSDE}) is a Gaussian white noise such that $\langle \eta(t)\eta(t') \rangle=\delta(t-t')$. We assume that a single sector boundary will only be able to move about a cell diameter $a$ per generation time $\tau_g$ due to cell rearrangement during division at the frontier.  Hence, $D_{\mathrm{2D}}\approx a^2/\tau_gR_0^2$, where  $R_0$ is the initial population radius. In addition,\ we assume that a green cell out-competing a red cell at the boundary can only shift the boundary by about a cell diameter.  Then, since a green cell out-competes a red cell at a rate proportional to the difference between their respective growth rates, we have $V_{\mathrm{2D}} \approx as/\tau_g R_0$.  A more detailed derivation of $D_{\mathrm{2D}}$ and $V_{\mathrm{2D}}$ is presented in Appendix C of \citet{MKNPRE}.
   Note that this random walk model is consistent with the dynamics of mutant sector sweeps described in Sec.~\ref{SPopGen}: if we set  $\Theta = 1$ (so that $t^*=R_0/v$) and look at vanishing genetic drift $(D_{\mathrm{2D}}=0)$,   Eq.~(\ref{eq:2dangleSDE}) yields the deterministic sector shape given by Eq.~(\ref{eq:2dlogspirals}) in the previous section with $ w_{\mathrm{2D}}= V_{\mathrm{2D}} t^*$.    

\subsection{Treadmilling   and Marginally Inflating Fronts $(\Theta \leq 1/2)$}

When $\Theta=0$, Eq.~(\ref{eq:2dangleSDE}) describes a drift-diffusion motion for the\ $\theta_{\mathrm{2D}}$ variable with a constant drift and diffusion constant.    At long times, there are only two possibilities for the fate of the sector:  It either eventually vanishes ($\theta_{\mathrm{2D}}=0$),   or it   wraps all the way around the population frontier ($\theta_{\mathrm{2D}}=2 \pi$).     The probability of the latter occurring is the survival probability. Using  standard first-passage techniques [see, e.g., Eq.~(2.3.8) in  \citet{redner}], we find that this survival probability for our model in Sec.~\ref{SModel} is
\begin{equation}
P_{\infty}^{\mathrm{treadmill}}(s,a,R_0)=\frac{1-e^{- V_{\mathrm{2D}} \theta_0/D_{\mathrm{2D}}}}{1-e^{-2 \pi  V_{\mathrm{2D}}/D_{\mathrm{2D}}} }\approx \frac{1-e^{-s N_0}}{1-e^{-2 \pi sR_0/a} }, \label{eq:nuttreadsurvp}
\end{equation}
 where $N_0 \approx R_0 \theta_0/a$ is the initial number of mutant cells at the frontier.  We also set $V_{\mathrm{2D}}/D_{\mathrm{2D}} =s R_0/a$ to describe expansions with compact frontiers that are a single cell wide.  Let us now consider a single initial mutated cell, so $N_0=1$.   In this case, Eq.~(\ref{eq:nuttreadsurvp}) resembles the classic Kimura formula for the fixation probability of a mutation with a selective advantage in a well-mixed population with initial frequency $f_0\approx a/2 \pi R_0$ and total population size $ N_{t}=2 \pi R_0/a$ (see, e.g., Eq.~8.8.3.13 in \citet{kimura}).   Indeed, as shown by \citet{maruyama1,maruyama2},  the survival probability of a mutation with a selective advantage is the same in well-mixed populations and at non-inflating population fronts.   When $s \rightarrow 0^+$, we find a non-zero survival probability.  Upon expanding Eq.~(\ref{eq:nuttreadsurvp})  for small $s$ (with $N_0=1$), we have
\begin{equation}
P_{\infty}^{\mathrm{treadmill}}(s,a,R_0) = \frac{a}{2 \pi R_0}+\frac{1}{2}\left[ 1- \frac{a}{2 \pi R_0}\right]s+\mathcal{O}(s^2) \qquad \qquad (s>0). \label{eq:nuttreadsurvpsmalls}
\end{equation}Note that when $s<0$ in Eq.~(\ref{eq:nuttreadsurvp}), the survival probability is exponentially suppressed, 
\begin{equation}  P_{\infty}^{\mathrm{treadmill}}(s,a,R_0) \approx e^{-2\pi |s|R_0/a}  \qquad \qquad (s<0)
 \end{equation}
for $R_0 \gg a /|s|$.

  For $0<\Theta \leq 1/2$, Eq.~(\ref{eq:2dangleSDE}) has a time-dependent drift and diffusion term. This makes the survival probability analysis more challenging.  However,  neutral mutations with $V_{\mathrm{2D}}=0$ may be treated exactly, as discussed in \citet{adi1,adi2,roughconformal,MKNPRE}.  In the limit of a large population front in which the mutant sector stays small ($\theta_{\mathrm{2D}} \ll 2\pi$), a neutral mutation  will  always eventually go extinct.   The $\Theta=\Theta_c=1/2$ case is marginal, and a neutral mutation is nearly able to survive.    We now  move to a discussion of the $\Theta > 1/2$ case, for which inflation produces a striking enhancement of the survival probability.  We will go back to the marginal case $\Theta = \Theta_c = 1/2$ when we treat three-dimensional expansions in Section~\ref{S3D}.

\subsection{Inflating fronts  $(\Theta>1/2)$}

When $\Theta>1/2$,   we may remove the time-dependence from the diffusion term in Eq.~(\ref{eq:2dangleSDE}) by changing variables from the time $t$ to a unitless time $\tau$ given by
\begin{equation}
\tau =  1- \left(1+\frac{t}{t^*} \right)^{1-2\Theta}  . \label{eq:taudef}
\end{equation}
  After this change of variables, Eq.~(\ref{eq:2dangleSDE}) becomes
\begin{equation}
\frac{d\theta_{\mathrm{2D}}}{d \tau} = \frac{t ^* V_{\mathrm{2D}}}{2\Theta-1}(1-\tau)^{\frac{\Theta}{2 \Theta-1}}+\sqrt{ \frac{ 2 t^* D_{\mathrm{2D}} }{2\Theta-1}} \, \eta(\tau), \label{eq:2dsectorboundary}
\end{equation}
where we again have a Gaussian white noise $\eta(\tau)$  such that $\langle \eta(\tau)\eta(\tau') \rangle=\delta(\tau-\tau')$.  Equation~(\ref{eq:2dsectorboundary}) describes a simple diffusive motion for the sector boundary in the conformal coordinate $\tau$, with a time-dependent bias.  Note that when $t=0$, then $\tau =0$, also.  However, if we send $t \rightarrow \infty$ in Eq.~(\ref{eq:taudef}), then $\tau \rightarrow 1$.  This means that the entire evolution of the sector, i.e., the interval $t \in [0,\infty)$, is compressed into the interval $\tau \in [0,1)$.    The consequences of this compression are explored in more detail in \citet{roughconformal, MKNPRE, adi1, adi2}.

The survival probability can be extracted from Eq.~(\ref{eq:2dsectorboundary}) by analyzing the first-passage probability that the sector angle $\theta_{\mathrm{2D}}$ vanishes  (see \citet{redner} for a review of first-passage processes).  Unlike the $\Theta=0$ case, we will have to work in the limit of large population fronts so that we can ignore large $\theta_{\mathrm{2D}} \approx 2 \pi$.  For $\Theta=1$, for example, we require that $R_0 \gg \sqrt{D_{\mathrm{2D}} t^*}$  and $\theta_{\mathrm{2D}}(t=\tau=0)=\theta_0 \ll 2\pi$. We will compare our results to simulations and to the exactly soluble $\Theta=0$ case to check the approximation.    We find a long-time survival probability that depends on two scaling variables $x_{\mathrm{2D}}$ and $\kappa_{\mathrm{2D}}$, given by
\begin{equation}
\begin{cases}
x_{\mathrm{2D}} = \dfrac{\theta_0}{\sqrt{D_{\mathrm{2D}}t^*}}=  \dfrac{R_0 \theta_0}{a} \sqrt{\dfrac{\tau_g}{ t^*}} \,\xrightarrow[\Theta = 1]{} \,\theta_0 \sqrt{\dfrac{R_0 v \tau_g}{ a^2}} \\[10pt]
\kappa_{\mathrm{2D}} =\dfrac{V_{\mathrm{2D}}t^*}{\sqrt{D_{\mathrm{2D}}t^*}}= s \sqrt{\dfrac{t^*}{ \tau_g}} \,\xrightarrow[\Theta = 1]{}  \,s \sqrt{\dfrac{R_0}{v \tau_g}}
\end{cases}, \label{eq:dimless2d}
\end{equation}
where we have again specialized to thin, compact population frontiers by setting  $V_{\mathrm{2D}}=a s/\tau_g R_0 $ and $D_{\mathrm{2D}} = a^2/\tau_gR_0^2 $. Note that we can only make the replacement $t^*=R_0/v$ in Eq.~(\ref{eq:dimless2d}) when $\Theta=1$.       

  The survival probability scaling function can be calculated using an adiabatic approximation.  Although the survival probability depends on many parameters, i.e. $P_{\mathrm{2D}}\equiv P_{\mathrm{2D}}(R_0, \theta_0, t^*, \tau_g, a , \Theta,t)$, we find that the full time-dependent solution for $\Theta>1/2$ in the adiabatic approximation depends on just  two parameter combinations [see Eq.~(\ref{eq:dimless2d})] and on the growth exponent $\Theta$:
\begin{align}
P_{\mathrm{2D}} & = P_{\mathrm{2D}}(x_{\mathrm{2D}}, \kappa_{\mathrm{2D}},\Theta, \tau) \nonumber \\ & \approx  1-  \int_0^{\tau} \mathrm{d} z \, \frac{x_{\mathrm{2D}}}{2} \sqrt{\frac{2\Theta-1}{ \pi  z^3}}  \exp\left\{-\frac{\left[ \kappa_{\mathrm{2D}}  z (1-z)^{\frac{\Theta}{1-2 \Theta}} +x_{\mathrm{2D}} (2\Theta-1)\right]^2 }{4z(2\Theta-1)} \right\}. \label{eq:2dsurvpscaling}
\end{align}
We can set $\tau = 1$ (i.e. $t \rightarrow \infty$) in Eq.~(\ref{eq:2dsurvpscaling}) to find the ultimate survival probability \\ $P_{\infty}(x_{\mathrm{2D}}, \kappa_{\mathrm{2D}},\Theta>1/2)$.   Although the integral does not appear to be expressible in terms of elementary functions, we may numerically integrate it for arbitrary parameter values. 

\begin{figure}[H]
\centering
\includegraphics[height=4.5in]{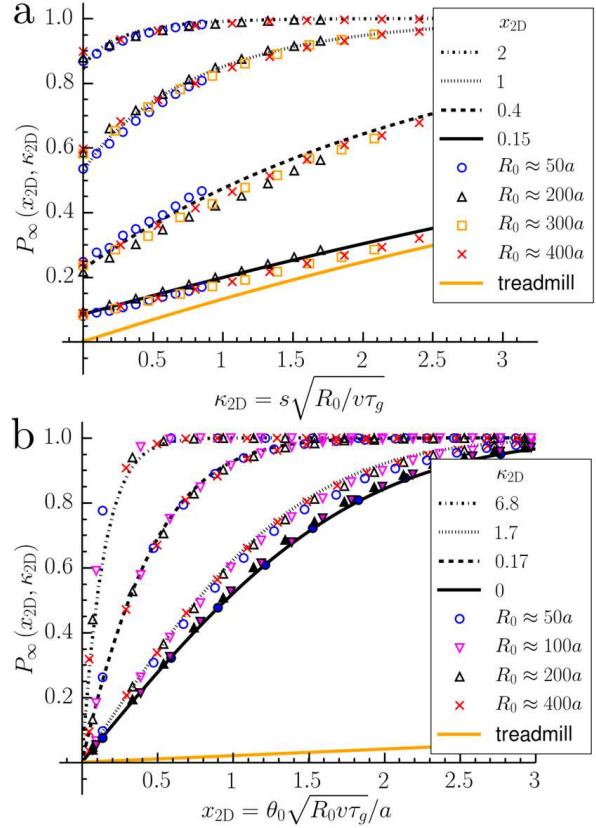}
\caption{\label{fig:circularsurvpscaling} The long-time ($t \rightarrow \infty$, $\tau \rightarrow 1$) survival probability scaling function $P_{\infty}$ of a green mutant sector [as in Fig.~\ref{fig:kappapics}(a)] for circular range expansions with $\Theta=1$, as a function of the two dimensionless variables $x_{\mathrm{2D}}$ and $\kappa_{\mathrm{2D}}$ displayed in Eq.~(\ref{eq:dimless2d}).  Clusters of contiguous mutants subtend an initial angle  $\theta_0$  along the population front of an otherwise all-red population with initial radius $R_0$. Different symbols lying on the same line correspond to data sets with different $R_0$.   In part (a),  $\theta_0$ is tuned to yield  a fixed $x_{\mathrm{2D}}$.   In part (b), $\theta_0$ is varied to find the probability at different $x_{\mathrm{2D}}$ at a fixed $\kappa_{\mathrm{2D}}$.   The solid symbols on the bottom curve in part (b) correspond to neutral range expansions with $s=0$. The lines show the analytic prediction using the adiabatic approximation  in Eq.~(\ref{eq:2dsurvpscaling}) for $\Theta=1$ (accurate for $0<s\ll 1$  and exact for $s=0$). In simulations (points), $P_{\infty}$ is estimated from the survival probability at a long (but finite) time $t \gtrsim 1200$ generations. Hence, we expect some deviation of the data from the analytic prediction, as the survival probability in simulations converges slowly to $P_{\infty}$ for large $R_0$ and small $s$. Solid orange lines show the survival probability for a treadmilling population front with $R_0=50a$ and a single initial mutant cell.  The orange line does go to a non-zero value as $\kappa_{\mathrm{2D}}=0$ in part (a), but it is indistinguishable from zero in the plot.   Note the greatly enhanced survival probability caused by inflation in part (a), as compared to treadmilling systems. }
\end{figure}

We now compare our analytic result, Eq.~(\ref{eq:2dsurvpscaling}), to simulations for $\Theta=1$.  Equation~(\ref{eq:2dsurvpscaling}) is numerically integrated for $\Theta=1$ to obtain the scaling function shown in Fig.~\ref{fig:circularsurvpscaling}.  To determine $\kappa_{\mathrm{2D}}$ and $x_{\mathrm{2D}}$ for our simulations (see Section~\ref{SModel}), we first set our length units so that $v \tau_g =1 $.    Then, the effective cell spacing $a \approx 0.9 v \tau_g$ is extracted from an independent measurement in \citet{MKNPRE}.  In addition, simulations of the deterministic motion of the sector yield $V_{\mathrm{2D}} = \omega_{\mathrm{2D}}/t^* \approx 1.2 as/\tau_g R_0$, as  mentioned in Sec.~\ref{SPopGen}.    Finally, setting $D_{\mathrm{2D}} \approx a^2/\tau_gR_0^2$, we find $\kappa_{\mathrm{2D}}$ and $x_{\mathrm{2D}}$ for our simulations using Eq.~(\ref{eq:dimless2d}).    This procedure is described in more detail in \citet{MKNPRE}.   The simulated survival probabilities are shown as data points in Fig.~\ref{fig:circularsurvpscaling}. We find that the simulation results match the predicted scaling form well.  We also show the treadmilling $\Theta=0$ result in Fig.~\ref{fig:circularsurvpscaling} for a fixed $R_0=50a$ to illustrate the large enhancement in the survival probability due to inflation.

As we approach a neutral mutation $\kappa_{\mathrm{2D}} \rightarrow 0$ in Fig.~\ref{fig:circularsurvpscaling}(a), we find a non-zero survival probability.  This is markedly different from the infinite flat front result in Eq.~(\ref{eq:linsurvp}), which vanishes as $s \rightarrow 0$. Indeed, for a single initial mutant cell at a circular front with initial radius $R_0$ ($x_{\mathrm{2D}} \approx \sqrt{a/ R_0}$), the limit is 
\begin{equation}
 P_{\infty}(x_{\mathrm{2D}}, \kappa_{\mathrm{2D}}=0)=   \mbox{erf} \left[\frac{x_{\mathrm{2D}}}{2} \right] \approx  \sqrt{\frac{a}{\pi R_0}}. \label{eq:survp2dnut}
\end{equation}
It is instructive to compare this result to the much smaller survival probability $a/(2\pi R_0)$ of a mutation occurring at the surface of a two-dimensional \textit{treadmilling} tumor.    The survival probability as $\kappa_{\mathrm{2D}} \rightarrow 0$ is much larger for an inflating rather than a treadmilling tumor,  as shown in Fig.~\ref{fig:circularsurvpscaling}(a). 
 Thus, for small selective advantages, inflation plays a much more important role in rescuing the mutation than one might expect from a naive analysis of a mutation sweeping a population with a finite size.     We will show in the next section that there is an analogous scaling function $P_{\mathrm{3D}}(x_{\mathrm{3D}}, \kappa_{\mathrm{3D}})$, with appropriate \textit{three-dimensional} analogues of the dimensionless scaling variables $x_{\mathrm{3D}}$ and $\kappa_{\mathrm{3D}}$.  We will again find that the effects of a finite population front size are less important than inflation at small $s$.  

\section{\label{S3D} Treadmilling and Expanding Populations in Three Dimensions}

\begin{figure}[H]
\centering
\includegraphics[width=2in]{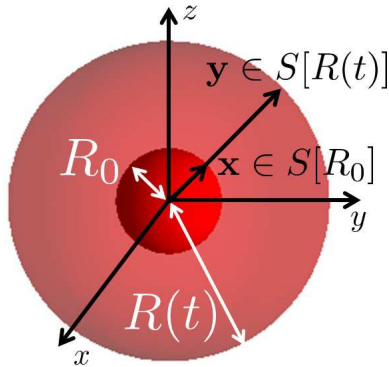}
\caption{\label{fig:coords}  Coordinate system for studying spherical expansions.  At time $t$ when the expansion has radius $R(t)$, we can map any point on the spherical population front, $\mathbf{y} \in S[R(t)]$, to a corresponding point $\mathbf{x} \in S[R_0]$ on the initial population front with radius $R_0$, along a radial line connecting $\mathbf{y}$ to the center.   }
\end{figure}

 A convenient coordinate system for three-dimensional range expansions is the position $\mathbf{ x} \in S(R_0)$ on the initial spherical population front $S(R_0)$.  Any positions $\mathbf{y} \in S[R(t)]$ on the spherical front $S[R(t)]$ of radius $R(t) = R_0(1+t/t^*)^{\Theta}$ at any time $t>0$ can be traced backward to a unique $\mathbf{x}=R_0 \mathbf{y}/R(t)$ on the initial front, as shown in Fig.~\ref{fig:coords}.  Thus, we can fully specify any position in the range expansion by a single position $\mathbf{x} \in S(R_0)$ and a time $t$.  We then interpret the evolution of the range expansion as a process acting on the original population front $S[R_0]$ with a superimposed inflation that locally dilates the front perpendicular to the local expansion direction.  The corresponding Langevin equation for the green cell fraction $f \equiv f(\mathbf{x},t)$ with this inflationary effect reads
\begin{equation}
\partial_t f = \frac{D}{(1+t/t^*)^{2\Theta}} \nabla_{\mathbf{x}}^2 f+s \tau_g^{-1} f(1-f)+\sqrt{\frac{2\Delta \ell^2 f\left[1-f \right]}{ \tau_g(1+t/t^*)^{2 \Theta}}} \, \xi(\mathbf{x},t), \label{eq:radlang}
\end{equation}
where $\xi(\mathbf{x},t)$ is again a Gaussian white noise induced by genetic drift [as in Eq.~(\ref{eq:linlang}), interpreted via the It\^o calculus] and $t^*$ is a characteristic time.    Equation~(\ref{eq:radlang}) is a non-linear stochastic partial differential equation with a spatial dilation.   Such equations are of interest not only in this biological context, but also in cosmology, interface growth, and other areas of condensed matter physics \citep{escudero}.
 In the case of linear inflation $(\Theta=1)$, we have $t^*=R_0/v$, where $v$ is the population front speed.  In this case, Eq.~(\ref{eq:radlang}) is a straightforward extension of the stochastic differential equation that describes \textit{circular}  range expansions \citep{MKNPRE,KorolevRMP}.  We now ignore the finite size of the sphere and extend our coordinate $\mathbf{x} \in S(R_0)$  to the infinite two-dimensional space $\mathbb{R}^2$.  This approximation will work as long as our sector sizes are small compared to the total surface area of the front.  We move to the field-theoretic representation of Eq.~(\ref{eq:radlang}), suitable for application of the methods of path integrals and statistical mechanics, and look for the ultimate survival probability $P_{\infty}$.  For a review of similar ideas from the perspective of population genetics, see \citet{bartonreview}. Readers unfamiliar with these formal developments may wish to pass on to our final conclusions in Sections~\ref{sec:treadmill3d} and \ref{sec:inflation3d}  which have been subjected to extensive numerical checks.
 
 Equation~(\ref{eq:radlang}) can be converted to a field-theoretic action by employing standard methods, pioneered over 40 years ago by  \citet{MSR}, and extended to  non-linear stochastic equations such as Eq.~(\ref{eq:linlang}) and Eq.~(\ref{eq:radlang}) by  \citet{janssenoriginal} and  \citet{dominicisoriginal}. We introduce a response function $\tilde{f}=\tilde{f}(\mathbf{x},t)$ to account for the genetic drift term in Eq.~(\ref{eq:radlang}).  This procedure yields the response functional
\begin{align}
\mathcal{J} [\tilde{f},f] & = \int \mathrm{d} t \,  \frac{\mathrm{d}^2 \mathbf{x}}{\ell^2} \, \left\{  \tilde{f} \left[  \frac{\partial f}{\partial t}-\frac{D}{(1+t/t^*)^{2\Theta}} \nabla^2 f -s \tau_g^{-1}f(1-f) \right. \right. \nonumber \\ & \qquad \qquad \qquad \qquad   \left. \left. {} - \frac{\Delta}{ \tau_g(1+t/t^*)^{2\Theta}} \, \tilde{f} f(1-f) \right]\right\}  , \label{eq:radaction1}
\end{align}
where $\ell^2$ is the area taken up by a single cell at the initial spherical frontier with radius $R_0$.  Equation~(\ref{eq:radaction1})  allows us to treat inflationary dynamics on the sphere using the methods of statistical mechanics. The genetic drift term (proportional to $\Delta$) is now on equal footing with the other contributions, making it easier to analyze than in the Langevin equation, Eq.~(\ref{eq:radlang}).  When such functionals are exponentiated, they determine the probability of particular path histories associated with the underlying stochastic differential equation \citep{bartonreview, MSR, janssenoriginal, dominicisoriginal}.

    The response functional in Eq.~(\ref{eq:radaction1}) can be used to calculate the survival probability for the initial patch of $N_0$ green cells.  As shown in \ref{appx:responsefunc},  a variational principle applied to the exponentiated response functional, Eq.~(\ref{eq:radaction1}), yields a spatially uniform, mean field approximation to the response function $\tilde{f}(\mathbf{x},t) \approx \tilde{f}(t)$.   With this mean field approximation, our long-term survival probability $P_{\infty}^{\Theta}$ for arbitrary growth exponents $\Theta$ [see Eqs.~(\ref{eq:PinfFTl1})-(\ref{eq:PinfFT}) in \ref{appx:responsefunc}] can be obtained by a limiting procedure,
\begin{equation}
P_{\infty}^{\Theta} \approx 1- \exp\left[{ \lim_{\zeta\rightarrow 0^+}\lim_{T \rightarrow \infty} N_0 \tilde{f}(t=0)} \right] ,\label{eq:meanfieldsurvp}
\end{equation}
where $\tilde{f}(t)$ satisfies
 \begin{equation}
  \frac{\partial \tilde{f}}{\partial t}=-  s  \tau_g^{-1}   \tilde{f} - \frac{\Delta}{  \tau_g(1+t/t^*)^{2\Theta}} \tilde{f}^2 +\zeta (1+t/t^*)^{2\Theta} \theta(T-t) , \label{eq:meanfield1}
\end{equation}
and where $\theta(x)$ is the step function ($\theta(x)=1$, $x>0$, $\theta(x)=0$, $x \leq 0$).

  Equation~(\ref{eq:meanfield1}) may be solved for $\tilde{f}(t=0)$ for  an arbitrary $\Theta \geq 0$.  This is done in the Appendices.   Substituting the solution for $\tilde{f}(t=0)$ into Eq.~(\ref{eq:meanfieldsurvp}) and taking the appropriate limits yields the survival probability
\begin{equation}
P_{\infty}^{\Theta} (N_0,s,\Delta,t^*,\tau_g) \approx 1-  \exp\left[- \, \frac{N_0 s  \theta(s) }{ \Delta } \,   \frac{e^{-s t^*/\tau_g}}{(st^*/\tau_g)^{2\Theta}\Gamma[1-2\Theta,st^*/\tau_g]}   \right], \label{eq:survpgeneralt}
\end{equation}
where $\Gamma(\alpha,x)$ is the incomplete gamma function \citep{ryzhik}.   We may now use Eq.~(\ref{eq:survpgeneralt}) to  identify  two key dimensionless parameters $x_{\mathrm{3D}}$ and $\kappa_{\mathrm{3D}}$:  
\begin{equation}
\begin{cases}
x_{\mathrm{3D}} =\dfrac{N_0 \tau_g }{\Delta t^*} \\[10pt]
\kappa_{\mathrm{3D}} =\dfrac{st^*}{\tau_g}
\end{cases}, \label{eq:dimless3d}
\end{equation}
where $N_0$ is the initial number of mutant cells.   Note the similarity between Eq.~(\ref{eq:dimless3d}) and the analogous dimensionless variables for circular inflation in Eq.~(\ref{eq:dimless2d}).
The survival probability only depends on these two parameters and the growth exponent $\Theta$: 
\begin{equation}
P_{\infty}^{\Theta}(x_{\mathrm{3D}},\kappa_{\mathrm{3D}}) \approx 1-  \exp\left[- \frac{x_{\mathrm{3D}}\theta(\kappa_{\mathrm{3D}} )\exp[-\kappa_{\mathrm{3D}}]}{\kappa_{\mathrm{3D}}^{2\Theta-1}\Gamma[1-2\Theta,\kappa_{\mathrm{3D}}]}   \right] . \label{eq:survpgeneralt2}
\end{equation}
 
 Equation~(\ref{eq:survpgeneralt2}) is the main result of our field theoretic approach.  We will check that it  approximates the survival probability well by comparing to simulations of treadmilling ($\Theta=0$) and linearly inflating ($\Theta=1$) population fronts.   First, however, we probe the properties of Eq.~(\ref{eq:survpgeneralt2}) by looking  at the survival probability for a neutral mutation (the   $\kappa_{\mathrm{3D}} \rightarrow 0^+$ limit).  A careful analysis of this limit yields
\begin{equation}
P_{\infty}^{\Theta}(x_{\mathrm{3D}},\kappa_{\mathrm{3D}}\rightarrow 0^+) \approx \begin{cases} 0 & \Theta < \frac{1}{2} \\ 1-   \exp\left[{- x_{\mathrm{3D}}   (2 \Theta-1)}  \right] & \Theta \geq \frac{1}{2} \end{cases}. \label{eq:nutgent}
\end{equation}
Equation~(\ref{eq:nutgent})  tells us that  $\Theta=1/2$ is an interesting ``marginal'' case for which inflation is nearly able to rescue a neutral mutation.   It is also possible to calculate the survival probability $P_{\infty}^{\mathrm{exp.}}$ of a neutral mutation on the frontier of an exponentially growing population $R(t) = R_0 e^{\lambda t}$.   If we substitute  $x_{\mathrm{3D}} =N_0 \tau_g /(\Delta t^*)$ in Eq.~(\ref{eq:nutgent}) and take $\Theta,t^* \rightarrow \infty$ with a constant ratio $\lambda = \Theta/t^*$, we find that $P_{\infty}^{\mathrm{exp.}}\approx 1-e^{- 2 N_0 \lambda \tau_g/\Delta}$.  We will consider the three cases $\Theta=0$, $\Theta=1$,  and $\Theta=1/2$ in more detail in the next subsections. 

\subsection{Treadmilling fronts $(\Theta=0)$ \label{sec:treadmill3d}}

When $\Theta=0$, the range expansion simply turns over cells at its surface, which we take for simplicity to be a one-cell-diameter thick spherical shell with radius $R_0$.   These dynamics are relevant for ``treadmilling'' tumors.  If we ignore the finite size of the shell  (a good approximation provided the mutant cluster is small, i.e., $R_0^2 \gg N_0 a^2$), the survival probability can be calculated from the solution to  Eq.~(\ref{eq:survpgeneralt2}) with $\Theta=0$.  We find a survival probability
\begin{align}
P_{\infty}^{\mathrm{treadmill}}[s,N_0,\Delta] & \approx \begin{cases}
1- \exp\left[ -\dfrac{N_0 s}{\Delta} \right] & s \geq 0 \\
0 & s<0
\end{cases} \label{eq:ftlinsurvp}
\end{align}  
for a very large population front $(R_0^2 \gg N_0 a^2)$. This result is consistent with the results of \citet{CDPFT} and with the generalized formula of \citet{doeringduality} in Eq.~(\ref{eq:linsurvp}).  We expect corrections for both positive and negative $s$ when the finite size of the population front is taken into account, as in the treadmilling circle result in Eq.~(\ref{eq:nuttreadsurvp}).    We note that our result, Eq.~(\ref{eq:ftlinsurvp}), is also valid for range expansions with large frontiers of a \textit{fixed} size that, instead of treadmilling, are growing in some direction by a single cell diameter $a$ per generation time $\tau_g$.  Since these fixed-size frontiers also do not experience inflation, we also expect them to obey our  $\Theta=0$ result, provided that the frontier is sufficiently large compared to the mutant cluster size.

\begin{figure}[H]
\centering
\includegraphics[width=3in]{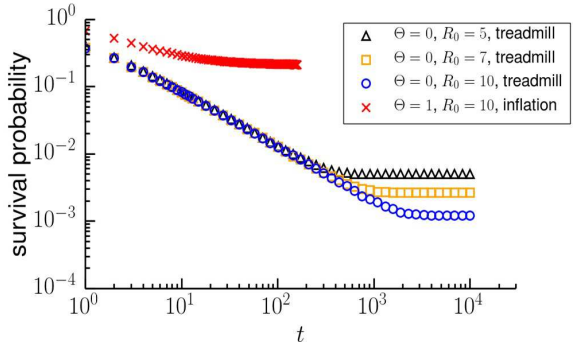}
\caption{\label{fig:infvstread} A comparison of treadmilling $(\Theta=0)$ and linearly inflating $(\Theta=1)$ expansion survival probabilities for single cell, neutral mutations at different initial radii $R_0$ (in units of the cell diameter $a$).  The survival probability for the treadmilling cases decays much faster with the time $t$ (measured in generations) and saturate at much smaller values. The plateaus at long times are consistent with $P_{\infty}^{\mathrm{treadmill}} \sim (a/R_0)^2$, as in Eq.~(\ref{eq:survptread3d}).  The survival probability for an inflating front saturates earlier at a much higher value.  The enhancement in the survival probability of inflating sectors for $R_0= 10 a$ is at least 100-fold.    }
\end{figure}

The neutral case $s=0$ is easy to analyze even for a finite size population front such as a treadmilling sphere.   We know from studies of voter model dynamics \citep{clusteringvoter, voter1} that the descendants of one of the initial cells will eventually sweep the entire population.  Hence, the survival probability of a single mutant cell out of a total of $N_{\mathrm{tot}}$ cells at the frontier is
\begin{equation}
P_{\infty}^{\mathrm{treadmill}}[s=0,N_0=1]=\frac{1}{N_{\mathrm{tot}}}\approx\frac{a^2 }{4 \pi R_0^2} \sim \left( \frac{a}{R_0} \right)^2. \label{eq:survptread3d}
\end{equation}
 We shall see in the next subsection that this treadmilling survival probability will be much smaller than the survival probability for an inflating tumor for large initial radii $R_0 \gg a$.  This expectation is already evident upon comparing a treadmilling and an inflating front in our simulations.  Note the dramatic qualitative difference in the survival probabilities with increasing time for $R_0/a=10$  in Fig.~\ref{fig:infvstread}.

 \subsection{Linearly inflating fronts $(\Theta=1)$ \label{sec:inflation3d}}

When $\Theta=1$, our characteristic time $t^*$ is the time to inflate to twice the initial radius $R_0$.  It satisfies $t^* = R_0/v$, where $v$ is the front speed.  Then, similarly to the $\Theta=1$ limit in the two-dimensional case [Eq.~(\ref{eq:dimless2d})] our dimensionless parameters in Eq.~(\ref{eq:dimless3d}) reduce to
\begin{equation}
\begin{cases}
x_{\mathrm{3D}} \xrightarrow[\Theta = 1]{} \dfrac{N_0 v \tau_g}{\Delta R_0} \\[10pt]
\kappa_{\mathrm{3D}} \xrightarrow[\Theta = 1]{}  \dfrac{s R_0}{v \tau_g}
\end{cases}. \label{eq:dimless3dt1}
\end{equation}
Using the $\Theta=1$ limit of the general result in Eq.~(\ref{eq:survpgeneralt2}), we find
\begin{align}
P_{\infty}(x_{\mathrm{3D}},\kappa_{\mathrm{3D}},\Theta=1) & \approx 1-  \exp\left[- \frac{x_{\mathrm{3D}}\theta(\kappa_{\mathrm{3D}} )e^{-\kappa_{\mathrm{3D}}}}{\kappa_{\mathrm{3D}}\Gamma[-1,\kappa_{\mathrm{3D}}]}   \right], \label{eq:spheresurvp} \\[6pt] &  \approx 1 - \exp\left[ - \frac{N_0(  \tau_g)^2\theta(s )e^{-s t^*/\tau_g}}{\Delta s (t^*)^2\Gamma[-1,s t^*/\tau_g]}\right]. \label{eq:survp3dscaling} 
\end{align}
  Note that $P_{\infty}=0$ when $s<0$. Hence, since $P_{\infty}>0$ for $s\rightarrow 0^+$, there is a jump discontinuity in the survival probability at the origin.  This discontinuity occurs because we have assumed that our inflating population front is effectively infinite relative to the size of the inflating sector.  Hence, when $s<0$, a mutant sector will always be able to shrink back deterministically to a small enough size that it can be extinguished via genetic drift.  This may happen very slowly, however, because the deterministic motion of the sector is logarithmic in time, as discussed in Section~\ref{SPopGen}.   Evaluation of the (small!) survival probability for negative $s$ for finite-sized fronts is a subject for future investigation (see also \ref{appx:negsel}).

It is instructive to study biologically relevant limits of Eq.~(\ref{eq:spheresurvp}).  For example, if we let $t^* \rightarrow \infty$ so that inflation becomes negligible, we recover the linear front $(\Theta=0)$ result, consistent with Eq.~(\ref{eq:linsurvp}) and Eq.~(\ref{eq:ftlinsurvp}):
\begin{equation}
P_{\infty}(x_{\mathrm{3D}},\kappa_{\mathrm{3D}} \gg 1) =1 - \exp\left[ -x_{\mathrm{3D}} \kappa_{\mathrm{3D}}+ \mathcal{O}(x_{\mathrm{3D}})\right]=1 - \exp \left[ -\frac{s N_0}{\Delta} \right].
\end{equation}
Another important limit is of course $s \rightarrow 0^+$ (so that $\kappa_{\mathrm{3D}} \rightarrow 0^+$).  We find
\begin{equation}
P_{\infty}(x_{\mathrm{3D}},\kappa_{\mathrm{3D}} = 0) \approx 1 - \exp \left[ -x_{\mathrm{3D}}\right] = 1 - \exp\left[ - \frac{N_0 \tau_g}{t^* \Delta}\right]. \label{eq:survpinf3dnut}
\end{equation}
Note that in our simulations, $t^*/\tau_g \approx R_0/a$.  Thus, for a single mutation $(N_0=1)$ and a small selection coefficient, we find, from Eq.~(\ref{eq:spheresurvp}), a linear increase in the survival probability as a function of $s>0$:
\begin{equation}
P_{\infty} \left(N_0=1,R_0,s\ll 1 \right)\approx  \frac{a}{\Delta R_0}-\ln \left[e^{\gamma_E} \dfrac{ R_0}{a}\,s\right] \dfrac{s}{\Delta}+\mathcal{O}(s^2 \ln^2 s ), \label{eq:survpsinglemut3d}
\end{equation}
where $\gamma_E \approx 0.577 $ is  Euler's constant.   In our simulations, the effective population size $N = \mathcal{O}(1)$, so  $\Delta \sim 1/N$ is on the order of unity as well.   The inverse scaling of the zeroeth order  term [$\mathcal{O}(s^0)$] in Eq.~(\ref{eq:survpsinglemut3d}) with the initial population radius is markedly different from the treadmilling result in Eq.~(\ref{eq:survptread3d}).  The inflation will rescue the mutation with much higher probability than the probability the mutation fixes at a treadmilling front due to the finite tumor size.   Also, the logarithm in the first order term [$\mathcal{O}(s)$] generates a diverging slope in $P_{\infty}$ as we send $s \rightarrow 0$.

\begin{figure}[H]
\centering
\includegraphics[height=4.4in]{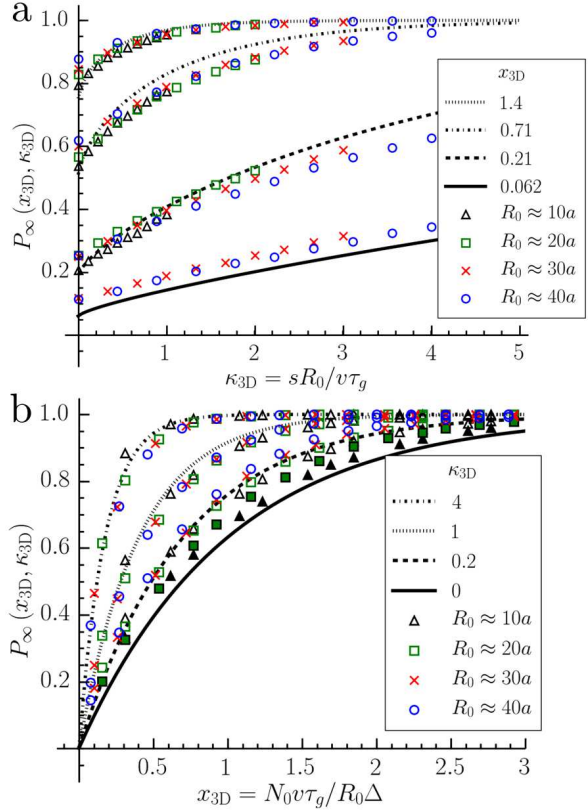}
\caption{\label{fig:sphericalsurvpscaling} The  survival probability scaling function of a mutant sector  for linearly inflating $(\Theta=1)$ spherical range expansions.  Clusters of mutants with $N_0$ initial cells start on the periphery of a population with initial radius $R_0$ and no other mutations. Different symbols lying on the same line correspond to data sets with different $N_0$ and $R_0$, but a fixed $x_{\mathrm{3D}}$ in part (a) and a fixed $\kappa_{\mathrm{3D}}$ in part (b).  All range expansions were evolved up to a total cell cluster with a 170 cell diameter radius. The solid symbols in part (a) correspond to neutral range expansions with $s=0$. The lines show the analytic prediction  in Eq.~(\ref{eq:survp3dscaling}). Note that at large values of $R_0$  and small values of $\kappa_{\mathrm{3D}}$ in part (a), the survival probability calculated in the simulations has not yet converged to the steady-state value, so the data points lie above the analytic prediction (lines).   }
\end{figure}

We check the approximation leading to our results with simulations in Fig.~\ref{fig:sphericalsurvpscaling}.  There are no fitting parameters, except the genetic drift strength $\Delta$. We find $\Delta \approx 0.6$, confirming our expectation that the genetic drift strength $\Delta$ is on the order of unity in our simulations.  The good agreement between the simulation results and the field theoretic solution justifies our approximation of an infinite population front. Indeed, from our scaling analysis in the previous section, we expect these particular finite size corrections to be small.  We are inherently limited by our finite simulation time, which allows us to only simulate clusters which have grown to radii of at most 170 cell diameters.

\subsection{Marginally inflating fronts $(\Theta=1/2)$}

\begin{figure}[H]
\centering
\includegraphics[height=2in]{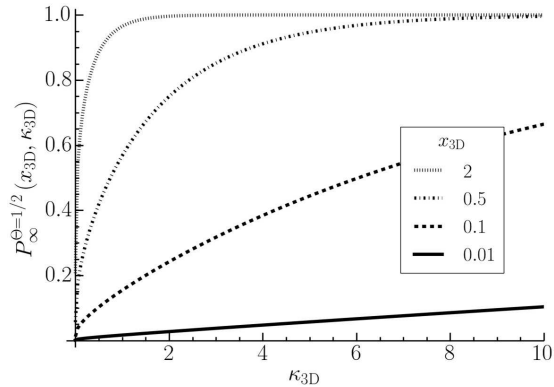}
\caption{\label{fig:marginalsurvp}  The survival probability for a mutation living on the surface of a spherical expansion with a $\Theta=1/2$ growth exponent [Eq.~(\ref{eq:radiusgrowth})].  The survival probability vanishes as $\kappa_{\mathrm{3D}} \rightarrow 0^+$, but  logarithmically slowly. }
\end{figure}

We now examine the marginal $\Theta=1/2$ case we identified previously in more detail.   We do not have simulation results for this case, but we may examine the analytic results.  Equation~(\ref{eq:survpgeneralt2}) yields a survival probability that reads
\begin{equation}
P_{\infty}^{\Theta=1/2}(x_{\mathrm{3D}},\kappa_{\mathrm{3D}}) =1- \exp \left[ -\frac{x_{\mathrm{3D}} \theta(\kappa_{\mathrm{3D}})}{ e^{\kappa_{\mathrm{3D}}}\Gamma(0,\kappa_{\mathrm{3D}})}\right],  \label{eq:marginalsurvp}
\end{equation}
where $\theta(\kappa_{\mathrm{3D}})=\theta(s)$  is again the step function. This result also reduces to the linear front solution when $t^* \rightarrow \infty$.  For small $\kappa_{\mathrm{3D}} $, we find the peculiar behavior
\begin{align}
P_{\infty}^{\Theta=1/2} (x_{\mathrm{3D}},\kappa_{\mathrm{3D}}\ll 1) & \approx 1- \exp \left[ -\frac{x_{\mathrm{3D}}\theta(\kappa_{\mathrm{3D}})}{\left|\ln(\kappa_{\mathrm{3D}}e^{\gamma_E})\right|}\right] \\
& \approx \frac{x_{\mathrm{3D}}\theta(\kappa_{\mathrm{3D}})}{\left|\ln(\kappa_{\mathrm{3D}}e^{\gamma_E}) \right|},
\end{align}
 where we assume $\kappa_{\mathrm{3D}} \ll e^{-x_{\mathrm{3D}}}$ in the second line. This probability vanishes as $\kappa_{\mathrm{3D}} \rightarrow 0^+$, just as we found in Eq.~(\ref{eq:nutgent}).  However, it does this logarithmically slowly.  We plot this unusual marginal scaling function in Fig.~\ref{fig:marginalsurvp}.

\section{\label{SConclusions}Conclusions} 

We showed how geometry plays a crucial role in the evolutionary dynamics of spherical range expansions, just as it does for circular ones.  We have calculated the long-time survival probability of a mutation at the spherical population frontier with either a fixed or growing radius  which takes the form $R(t) \sim (t/t^*)^{\Theta}$ with arbitrary growth exponent $\Theta>0$ for $t$ large compared to a characteristic time $t^*$.  Our primary focus has been on linearly inflating ($\Theta=1$) and treadmilling ($\Theta=0$) populations, both of which are relevant for tumor evolution. For linearly inflating fronts, we have $R(t) = R_0(1+vt/R_0)$, where $v$ is the front velocity, and $t^*=R_0/v$ is the time required for the front to be affected by the inflationary geometry. Inflating population fronts yield an enhanced survival probability of mutations at range expansion frontiers in two and three dimensions.   We have shown how this survival probability scales with the population front radius $R_0$ when an isolated mutation occurs.  The long time survival probability $P_{\infty}$ of a mutation decreases with increasing $R_0$ as $P_{\infty}^{\Theta=1} \sim \sqrt{a/R_0}$ in two dimensions and $P_{\infty}^{\Theta=1} \sim a/R_0$ in three dimensions for linearly inflating range expansions, where $a$ is a typical cell diameter.  The corresponding probabilities for \textit{treadmilling} populations are  $P_{\infty}^{\Theta=0} \sim a/R_0$ and $P_{\infty}^{\Theta=0} \sim (a/R_0)^2$ for two and three dimensions, respectively.  Hence, inflation rescues mutations with a much higher probability than might be expected in populations with a fixed front size.  For example, for  spherical populations with thin population fronts and initial radii of ten cell diameters,  $R_0=10a$, our analytic results for treadmilling populations and linearly inflating range expansions in Eq.~(\ref{eq:survptread3d}) and Eq.~(\ref{eq:survpinf3dnut}), respectively, yield a 100-fold enhancement for inflating expansions: $P_{\infty}^{\Theta=1}/P_{\infty}^{\Theta=0} \approx 100$.  We also verified this remarkably large enhancement in simulations, as shown in Fig.~\ref{fig:infvstread}.

 We have also gone beyond simple scaling and calculated the scaling function for the survival probability in the infinite time limit, $P_{\infty}^{\Theta}$,  for spherical expansions.  We applied field-theoretic techniques to calculate $P_{\infty}^{\Theta}$ for arbitrary $\Theta$.     On the other hand, using random walk theory, we have recaptured some previous results \citep{MKNPRE} for the survival probability in circular range expansions, and generalized the results  to an arbitrary  $\Theta$ [see Eq.~(\ref{eq:radiusgrowth})].  Using these two different techniques, we found that the scaling functions depend on key dimensionless parameters: a rescaled selection coefficient, $\kappa_{\mathrm{2D},\mathrm{3D}}$,  and a rescaled initial number of mutant cells, $x_{\mathrm{2D},\mathrm{3D}}$ in two (2D) and three (3D) dimensions.  We also looked at the fascinating case of marginal inflation, $\Theta=1/2$, where the population radius increase is almost able to rescue a neutral mutation from extinction via genetic drift. It would be interesting to apply the field-theoretic techniques described here to different evolutionary dynamics.  For example, we may include more complicated, ``mutualistic'' interactions between cell types at tumor surfaces, which may occur when certain cell strains promote the growth of surrounding tumor cells \citep{Marusyk}.

 The survival probability of a mutation in a three-dimensional range expansion captured by  Eq.~(\ref{eq:survpgeneralt2}) has many biological applications.  For example, we may evaluate a mutation survival probability  for tumors, which have growth exponents that can decrease from $\Theta=1$ to $\Theta=0$ as the tumor approaches a ``treadmilling'' regime.  Of course, $\Theta$ in such tumors is time-dependent, and our results will hold only when $\Theta$ varies  slowly compared to the time it takes for the mutation's fate to be decided.   Then, if this variation in $\Theta$ is slow, we could substitute the appropriate $\Theta$ into Eq.~(\ref{eq:survpgeneralt2}) to find a mutation survival probability at various tumor growth stages.   A precise calculation of how slow the variation of $\Theta$ has to be and how the survival probability converges to its infinite-time limit $P_{\infty}^{\Theta}$ for arbitrary $\Theta$ would be an interesting subject for future work.

  Our results may be tested experimentally in microbial populations, or, possibly, in various human cell lines \citep{breastcancer}.  For example,  a spherical  range expansion with $\Theta=1$ might be created by growing yeast cell colonies in soft agar.  Conversely, growing yeast cell \textit{pillars} with a fixed cross-section, as described in \citet{vulin}, would generate an expansion with a fixed-size frontier and no inflation.  Such expansions would obey our $\Theta=0$ results in Eq.~(\ref{eq:ftlinsurvp}) for sufficiently large frontier sizes. Controlled cell cluster growth and the approach to a treadmilling regime may also be realized, for example, by confining tumor spheroids in hollow elastic microspheres \citep{Alessandri}.   Again, we predict a strongly enhanced survival probability of mutations in  the spherical expansions.  This result could be checked using a fluorescing yeast strain, for example.   If a circular
or spherical range expansion has a negligible surface tension at the frontier, then the population front 
will roughen over time as the cells stochastically divide and push each other into virgin territory. This
roughening can affect the evolutionary dynamics.  Thus, another interesting avenue for future research, initiated by \citet{andersonunpub},  is to calculate scaling functions for the survival probability of mutations on a rough population front with a non-zero selective advantage $s > 0$.  It would also be interesting to study the $s<0$ regime in more detail.  Since the convergence of the survival probability to its infinite time value is so slow in this regime (see \ref{appx:negsel}), a time-dependent theory would be necessary to find the survival probability even at very long times.

\section{\label{SAcknowledgements}Acknowledgements} 

 We thank James Glazier, Oscar Hallatschek, and Kirill Korolev  for helpful discussions. One of us (DRN) would like to acknowledge helpful conversations about tumor growth with Suzanne Gaudet. This work
was supported in part by the National Science Foundation (NSF) through grant DMR-1306367, and by the Harvard Materials Research Science and Engineering Center
 (grant DMR-1420570).  Parts of this work were conducted during the workshop ``Cooperation and the Evolution of Multicellularity'' at the Kavli Institute for Theoretical
Physics at Santa Barbara, supported by the NSF through
grant PHY11-25915. Computational resources were provided by the Harvard Odyssey Research Computing group and the Center for Nanoscale Systems (CNS), a
member of the National Nanotechnology Infrastructure
Network (NSF grant ECS-0335765). CNS is part of Harvard University.
MOL acknowledges support from NSF grant  DMR-1262047.

\appendix

\section{\label{appx:responsefunc} Response functional formalism}

 The response functional formalism allows us to calculate moments of the coarse-grained green cell fraction $f(\mathbf{x},t)$ (with $\mathbf{x}$ some position along the front at time $t$), such as $\langle f(\mathbf{x},t)\rangle$, $\langle f(\mathbf{x},t) f(\mathbf{x}',t') \rangle$, etc., where the averages are over all possible realizations of the noise $\xi(\mathbf{x},t)$ in the Langevin equation (\ref{eq:radlang}) obeyed by $f(\mathbf{x},t)$.    Schematically, then, for any functional $\mathcal{G}[f]$ of $f \equiv f(\mathbf{x},t)$, we may take an average $\langle \mathcal{G}[f] \rangle = \int \mathcal{D} \xi \, \mathcal{G}[f]P[\xi]$, where we integrate over all possible noise instances $\xi(\mathbf{x},t)$, each of which has a probability weight
\begin{equation}
P[\xi] \propto \exp \left[ -\int \mathrm{d}^2 \mathbf{x} \, \mathrm{d} t \, \, \frac{ \xi(\mathbf{x},t)^2}{2 } \right].
\end{equation}    
We perform the integration over $\xi$ by introducing a functional Dirac delta function.  This allows us to change variables from $\xi(\mathbf{x},t)$ to $f(\mathbf{x},t)$, ensuring that $f(\mathbf{x},t)$ obeys Eq.~(\ref{eq:radlang}), which we can write generally as
\begin{equation}
\partial_t f(\mathbf{x},t)=F[f(\mathbf{x},t)]+\sqrt{W[f(\mathbf{x},t)]}\xi(\mathbf{x},t),
\end{equation}
    where $F[\ldots]$ is the force term and $\sqrt{W[\ldots]}$ is the noise strength.  Both of these are easily identifiable in Eq.~(\ref{eq:radlang}). The integration over $\xi$  then becomes a Gaussian integral as follows:
\begin{align}
\langle \mathcal{G}[f(\mathbf{x},t)] \rangle & = \int \mathcal{D} \xi \, \mathcal{D} f \, \mathcal{G}[f] \,\delta\left(\partial_t f - F[f]-\sqrt{W[f]}\,\xi\right)\,P[\xi] \nonumber \\ & = \int \mathcal{D} \xi \,\mathcal{D}f\,\mathcal{D}[i\tilde{f}]\,\mathcal{G}[f] \nonumber \\ & \qquad \times  \exp \left[ -\int \mathrm{d}^2 \mathbf{x}\,\mathrm{d}t\, \frac{\tilde{f}}{\ell^2} \left(\partial_t f - F[f]- \sqrt{W[f]}\, \xi \right) -\int \mathrm{d}^2 \mathbf{x} \, \mathrm{d} t \, \, \frac{ \xi(\mathbf{x},t)^2}{2 }\right] \nonumber \\ 
& =\int \mathcal{D}f\, \mathcal{D} [i\tilde{f}] \, \mathcal{G}[f] \exp \left[ -\int \mathrm{d}^2 \mathbf{x} \, \mathrm{d}t \, \frac{\tilde{f}}{\ell^2}\left( \partial_t f - F[f]-\frac{ \tilde{f}W[f]}{2 \ell^2}\right) \right]  \label{eq:averagesFT} \\
& \equiv \int \mathcal{D} f \, \mathcal{D} [i \tilde{f}] \, \mathcal{G}[f] \, e^{- \mathcal{J}[f,\tilde{f}]}, \label{eq:averagesFT2}
\end{align} 
where we have rewritten the functional Dirac delta function $\delta(\ldots)$ as a functional Fourier transform by introducing a function $\tilde{f} = \tilde{f}(\mathbf{x},t)$ (the analogue of the wave-number $\mathbf{k}$ for the usual Fourier transform) that we integrate over the imaginary axis for every $\mathbf{x}$ and $t$. We choose the normalization of $\tilde{f}$, $1/\ell^2$, to make $\tilde{f}$ dimensionless, just like $f$.  The function $\tilde{f}(\mathbf{x},t)$ is often called a response field.  (Note that we have neglected the Jacobian associated with changing variables in the delta function and we have absorbed any normalization constants in the functional measures $\mathcal{D} f$, $\mathcal{D} [i \tilde{f}]$.  For more details and subtleties, we refer to the  book by \citet{tauberbook}.)  We see that the force and noise terms in Eq.~(\ref{eq:averagesFT}) are now on equal footing.  Indeed, we can now interpret any average $\langle \ldots \rangle$ over instances of the noise $\xi(\mathbf{x},t)$ as averages weighted by a Boltzmann-like factor $e^{- \mathcal{J}[\tilde{f},f]}$, where $\mathcal{J}[\tilde{f},f]$ is given by Eq.~(\ref{eq:radaction1}) and we integrate over all possible $f(\mathbf{x},t)$ and $\tilde{f}(\mathbf{x},t)$.

 We now want to use the response function formalism to  determine the survival probability.  We follow Janssen's analysis \citep{CDPFT} and consider first the probability of forming an $N$-cell sector of mutants from an $N_0$-particle initial one in some time $T\gg \tau_g$.    We  introduce an initial fraction $f_0(\mathbf{x})$ of green cells by adding an additional term to the field theoretic action $\mathcal{J}[f,\tilde{f}]$:
 \begin{equation}
 -\int \frac{\mathrm{d}^2 \mathbf{x}}{\ell^2} \, f_0(\mathbf{x}) \, \tilde{f}(\mathbf{x},t=0) \approx- \int \mathrm{d}^2 \mathbf{x} \,N_0 \delta(\mathbf{x}) \, \tilde{f}(\mathbf{x},t=0) = -N_0 \tilde{f}(\mathbf{0},0), \label{eq:initialterm}
 \end{equation}
  where the first equality assumes a concentrated aggregation of $N_0$ cells near the origin $\mathbf{x}=0$ at time $t=0$.  

   We can calculate the probability by counting up the fraction of noise histories  $\xi(\mathbf{x},t)$ that yield a cluster of $N$ cells.  Hence, using the functional integration technique in Eq.~(\ref{eq:averagesFT2}), we find
\begin{align}
P_N(T) & \approx \int \mathcal{D} f \, \mathcal{D} \tilde{f} \, \delta\left[N - \int \frac{\mathrm{d}^2 \mathbf{x}}{\ell^2}\, \mathrm{d}t\, \theta(T-t) \left( 1+\frac{t}{t^*}\right)^{2 \Theta}f(\mathbf{x},t) \right] \, e^{- \mathcal{J}[\tilde{f},f]+N_0 \tilde{f}(\mathbf{0},0)}, \label{eq:PN}
\end{align}
where $\theta(x)$ is the step function, and we have explicitly included the initial condition term in the exponential (see Eq.~(\ref{eq:initialterm})).  The Dirac delta function  $\delta(\ldots)$ in Eq.~(\ref{eq:PN}) ensures that we only count the noise histories that yield a cluster of $N$ cells. These probabilities $P_N(T)$ can be used to find the survival probability via a generating function or discrete Laplace transform
\begin{align}
G(\zeta,T) &\equiv \sum_{N=0}^{\infty} P_N (T)e^{-\zeta N} \\ & \approx \int \mathrm{d} N  \int \mathcal{D} f \, \mathcal{D} \tilde{f} \,  \delta\left[N - \int \frac{\mathrm{d}^2 \mathbf{x}}{\ell^2}\, \mathrm{d}t\,\left( 1+\frac{t}{t^*}\right)^{2 \Theta} \theta(T-t) f(\mathbf{x},t) \right]  e^{- \mathcal{J}[\tilde{f},f]+N_0 \tilde{f}(\mathbf{0},0)-\zeta N}  \\ & =  \int \mathcal{D} f \, \mathcal{D} \tilde{f} \,   e^{- \mathcal{J}[\tilde{f},f]+N_0 \tilde{f}(\mathbf{0},0)-\zeta \int \frac{\mathrm{d}^2 \mathbf{x}}{\ell^2}\, \mathrm{d}t\,\left( 1+\frac{t}{t^*}\right)^{2 \Theta} \theta(T-t) f(\mathbf{x},t)} .
\end{align} 
The generating function $G(\zeta,T)$ is related to the long-time survival probability $P_{\infty}$ of the mutant cluster, which turns out to be given by: \begin{align}
P_{\infty}& =1- \lim_{\zeta \rightarrow 0^+} \lim_{T \rightarrow \infty} G(\zeta) \label{eq:PinfFTl1}\\ & \approx1- \lim_{\zeta \rightarrow 0^+} \lim_{T \rightarrow \infty}  \int \mathcal{D} f \, \mathcal{D} \tilde{f} \,   e^{- \mathcal{J}[\tilde{f},f]+N_0 \tilde{f}(\mathbf{0},0)-\zeta \int \frac{\mathrm{d}^2 \mathbf{x}}{\ell^2}\, \mathrm{d}t\,  \left( 1+\frac{t}{t^*}\right)^{2 \Theta}\theta(T-t)f(\mathbf{x},t)} \label{eq:PinfFTl2} \\
P_{\infty}& \approx 1- \lim_{\zeta \rightarrow0^+} \lim_{T \rightarrow \infty} \left\langle e^{N_0 \tilde{f}(\mathbf{0},0)} \right\rangle_{\mathcal{J}[\tilde{f},f]+\zeta  \int \frac{\mathrm{d}^2 \mathbf{x}}{\ell^2}\, \mathrm{d}t\, \left( 1+\frac{t}{t^*}\right)^{2 \Theta} \theta(T-t)f(\mathbf{x},t)},  \label{eq:PinfFT}
\end{align}
where the average is performed using the exponentiated field theoretic action in Eq.~(\ref{eq:radaction1}), shifted by the step function term: $\mathcal{J}[\tilde{f},f] \rightarrow \mathcal{J}[\tilde{f},f]+\zeta  \int \frac{\mathrm{d}^2 \mathbf{x}}{\ell^2}\, \mathrm{d}t\,  \left( 1+\frac{t}{t^*}\right)^{2 \Theta}\theta(T-t)f(\mathbf{x},t)$.

For a mutant survival probability on an infinite, flat front [$\Theta=0$ in Eq.~(\ref{eq:radlang})], it is known that a mean-field approximation (up to logarithmic corrections) works around the special ``voter model'' case of $s=0$ with one time dimension and two spatial dimensions.  Here, we assume mean-field theory also works for $0 < s \lesssim 1$ and in the presence of inflation [$\Theta>0$ in Eq.~(\ref{eq:radlang})].  We will check our mean-field approximation by comparing the results to simulations and find that it works remarkably well.   Within the mean-field approximation, the average in Eq.~(\ref{eq:PinfFT}) can be evaluated by substituting in the spatially uniform, saddle-point approximation for $\tilde{f}(t=0)\equiv\tilde{f}(\mathbf{0},0)$.  The saddle point equations, obtained by varying the shifted action $\mathcal{J}[f,\tilde{f}]$  with respect to $\tilde{f}$ and $f$ are, respectively, $f=0$ and Eq.~(\ref{eq:meanfield1}) in the main text.  We can rewrite the latter equation by introducing a new  time variable $\bar{t} \equiv t/T$ so that $\bar{t} \in [0,1]$. Then, incorporating the step function in Eq.~(\ref{eq:meanfield1}) into an appropriate boundary condition, we find:
\begin{equation}
\partial_{\bar{t}}\tilde{f}(\bar{t}) =- s  \tau_g^{-1} T\tilde{f}(\bar{t})- \frac{\Delta T [\tilde{f}(\bar{t})]^2}{ \tau_g(1+\bar{t} T/t^*)^{2\Theta}}+\zeta\left( 1+\bar{t} T/t^*\right)^{2 \Theta} T \mbox{\quad with \quad} \tilde{f}(1)=0. \label{eq:meanfield2}
\end{equation}
This equation can be solved for $\tilde{f}(t=0)$ in the $T \rightarrow \infty$ and $\zeta \rightarrow 0^+$ limit for various growth exponents $\Theta$.  We will do this in \ref{appx:survp}.

\section{\label{appx:survp} Survival probability calculation in three dimensions}

We want to solve Eq.~(\ref{eq:meanfield2}) for various $\Theta$.   To emphasize that we are looking for solutions at arbitrary $\Theta$, let us introduce a subscript to our response function: $\tilde{f}_{\Theta}(\bar{t})$. Recall that Eq.~(\ref{eq:meanfield2}) describes a time history of $\tilde{f}_{\Theta}(\bar{t})$ in the range $0 \leq \bar{t} \leq 1$.  Equation~(\ref{eq:meanfield2}) is a Ricatti equation with time-dependent coefficients \citep{ricattibook}. We can solve it by introducing a function $u_{\Theta}(\bar{t})$ such that  $\tilde{f}_{\Theta} (\bar{t})=(1+\bar{t} T/t^*)^{2\Theta}  \partial_{\bar{t}} u_{\Theta}(\bar{t})/[u_{\Theta} (\bar{t})\Delta T]$. Equation~(\ref{eq:meanfield2}) is then transformed into a linear second-order differential equation that reads
\begin{align}
\frac{d^2 u_{\Theta}}{d\bar{t}^2}+ \left( sT+\frac{2 \Theta T/t^* }{1+\bar{t}T/t^*}\right) \,\frac{du_{\Theta}}{d\bar{t}}- \Delta\zeta T^2\,u_{\Theta}=0, \label{eq:ualphaDE}
\end{align}
where we have also set $\tau_g=1$ without loss of generality.  To restore the generation time, we simply have to divide both $s$ and $\Delta$ by $\tau_g$.  Then, upon changing variables to $z \equiv s(t^*+T\bar{t}) \nu$ (where $\nu^2 \equiv 1+4 \Delta \zeta/s^2$) and introducing a new function $w_{\Theta}(z)\equiv e^{z(1+\nu^{-1})/2} u_{\Theta}(z)$, we find the confluent hypergeometric differential equation
\begin{equation}
z \, \frac{d^2 w_{\Theta}}{dz^2} +(2 \Theta-z ) \frac{dw_{\Theta}}{dz}-\left( 1+\nu^{-1}\right) \Theta w_{\Theta}(z)=0.
\end{equation}
 The general solution for $w_{\Theta}(z)$ is given by a linear combination of special functions \citep{ryzhik},
\begin{equation}
w_{\Theta}(z) =A_1 \,\, {}_1F_1 \left[(1+\nu^{-1})\Theta;2 \Theta;z  \right]+A_2\, U \left[(1+\nu^{-1})\Theta,2 \Theta,z  \right],
\end{equation}
where $A_1$ and $A_2$ are arbitrary constants, and  ${}_1F_1[\alpha;\beta;z]$ and  $U[\alpha,\beta,z]$   are the confluent hypergeometric functions of the first and second kind, respectively. Upon substituting this general solution into the expression for  $ \tilde{f}_{\Theta}(z)$, we find 
\begin{align}
\tilde{f}_{\Theta}(z)
& =- \, \frac{s z^{2 \Theta}  }{ \Delta (s \nu t^*)^{2 \Theta} } \,  \left\{ \frac{1+\nu^{-1}}{2} \right. \nonumber \\ & \qquad \left.   {}+ \left[1+\nu^{-1}\right]\frac{ \Theta  U\left[1+(1+\nu^{-1})\Theta,1+2\Theta,z \right]-\frac{B}{2}\, {}_1F_1 \left[1+(1+\nu^{-1})\Theta;1+2\Theta;z \right] }{ U \left[(1+\nu^{-1})\Theta,2\Theta,z \right]+B\,{}_1F_1 \left[(1+\nu^{-1})\Theta;2\Theta;z \right] }  \right\},
\end{align}
where we will now fix  $B \equiv A_1/A_2$ by applying the boundary condition $\tilde{f}_{\Theta=1}[z=s \nu (t^*+T)]=1$.  In the $T \gg t^*$ limit, we have $s\nu (t^*+T) \approx s \nu T$.  Our boundary condition, after taking the $\zeta \rightarrow 0^+$ ($\nu\rightarrow 1$) limit, is 
\begin{align}
\tilde{f}_{\Theta}(z=sT) & =- \frac{s }{\Delta}\left( \frac{T}{t^*} \right)^{2 \Theta} \left[1 +\frac{ 2\Theta  U\left[1+2\Theta,1+2\Theta,sT \right]-B\, {}_1F_1 \left[1+2\Theta;1+2\Theta;sT \right] }{ U \left[2\Theta,2\Theta,sT \right]+B\,{}_1F_1 \left[2\Theta;2\Theta;sT\right] } \right]\nonumber \\ & =-\frac{s }{\Delta} \left( \frac{T}{t^*} \right)^{2 \Theta}\left[1 + \frac{ 2\Theta  U\left[1+2\Theta,1+2\Theta,sT \right]e^{-sT}-B}{ U \left[2\Theta,2\Theta,sT \right]e^{-sT}+B  } \right]=1.
\end{align}
 The solution for $B$ is
\begin{align}
B =-\left\{ \frac{2\Theta s }{\Delta}\,  U\left[1+2\Theta,1+2\Theta,sT \right]+\left[\left( \frac{t^*}{T} \right)^{2 \Theta}+ \frac{s}{\Delta}\right]U\left[2\Theta,2\Theta,sT  \right] \right\}\left( \frac{T}{t^*} \right)^{2 \Theta}e^{-sT}.
\end{align}
In the long time $T \rightarrow \infty$ limit, we will have $B \rightarrow 0$ if $s>0$ and $|B| \rightarrow \infty$ if $s<0$.  Hence, we treat these cases separately. We will find a jump discontinuity at $s=0$ for $\Theta>1/2$, as discussed in the main text.  To find the survival probability, we must evaluate $\tilde{f}_{\Theta}(\bar{t})$ at $\bar{t}=0$ (or $z=s \nu t^*$).  The calculation for $s>0$ in the $T \rightarrow \infty$ limit leads to
\begin{align}
\tilde{f}_{\Theta}(z=st^*)& =- \, \frac{s   }{ \Delta } \,  \left\{ 1 +\frac{ 2\Theta  U\left[1+2\Theta,1+2\Theta,s t^*\right]e^{-st^*}-B }{ U \left[2\Theta,2\Theta,s t^* \right]e^{-st^*}+B }  \right\} \nonumber \\
& = - \, \frac{s   }{ \Delta } \,  \left\{ 1 +\frac{ 2\Theta  \Gamma\left[-2\Theta,s t^*\right] }{ \Gamma \left[1-2\Theta,s t^* \right] }  \right\} =- \, \frac{s   }{ \Delta } \frac{e^{-s t^*}(st^*)^{-2\Theta}}{\Gamma[1-2\Theta,st^*]}, 
\end{align}
where $\Gamma(\alpha, x)$ is the incomplete gamma function \citep{ryzhik}. For $s<0$, we have $\tilde{f}_{\Theta}(st^*) = 0$.   We may now substitute in various values of $\Theta$ to get our survival probabilities.

\section{\label{appx:negsel} Survival probability for $s<0$}

\begin{figure}[H]
\centering
\includegraphics[height=4in]{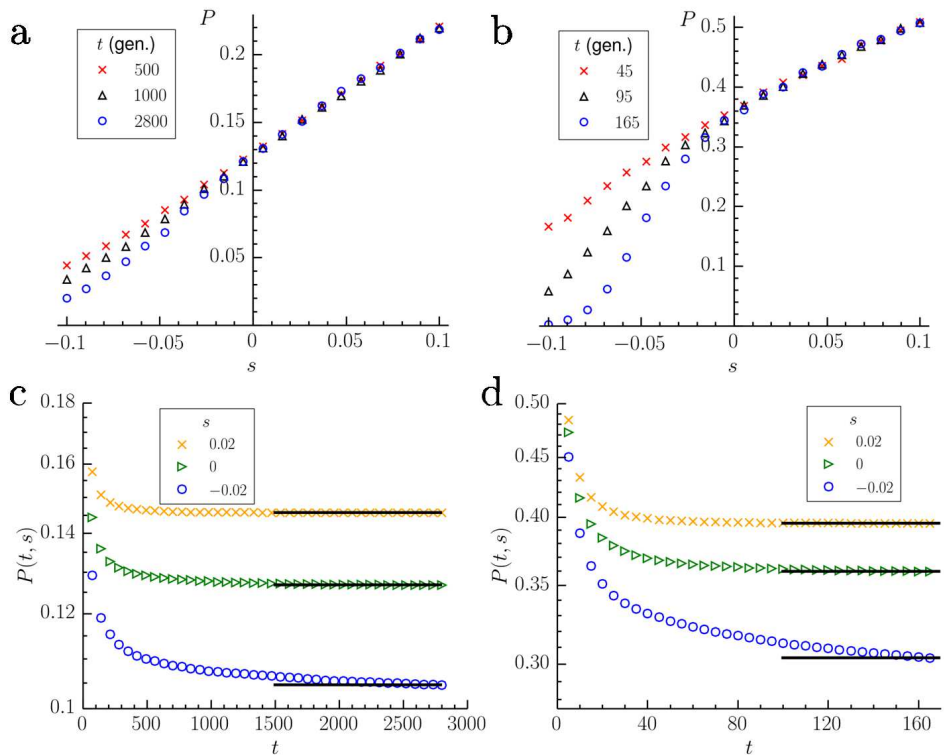}
\caption{\label{fig:negsel} The  survival probability $P\equiv P(t,s)$ at time $t$ generations  of a cluster (with selection parameter $s$) formed from a single neutral green cell at (a,c) a circular frontier with initial radius $R_0=20$ cell diameters  and (b,d) a spherical frontier with  $R_0=5$ cell diameters.  We plot $P$ as a function of $s$ for various $t$  in (a,b) and as a function of $t$ for a few $s$ values in (c,d) (with a logarithmic vertical scale). When $s<0$ in (a,b), the survival probability has not yet converged and continues to decrease with increasing $t$.  Note that in the $s=-0.02$ case in (c,d), the probability continues to decrease as we increase $t$, indicating that it has not yet converged to a long-time value.   The horizontal black lines denote the survival probability values at the longest time simulated [2800 generations in (c) and 170 generations in (d)]. Our theory predicts the probability will eventually decay to a vanishingly small (compared to, say, the limiting probability for $s=0^+$) residual  probability due to the very small chance that sectors wrap around the entire population. }
\end{figure}

Our theory predicts that a mutation with a selective disadvantage $s<0$ occurring at the frontier of a circular or spherical population only survives if genetic drift allows it to wrap all the way around the population.  Otherwise the deterministic dynamics will collapse the mutant sector and $P(t \rightarrow \infty,s<0) \equiv P_{\infty}(s<0)=0$.  However, the convergence to this limiting survival probability is very slow, due to the deterministic dynamics (logarithmic spiral shapes, etc.) described in the main text.  To show this, we plot the survival probability $P(t,s)$ at various times $t$ for circular and spherical frontiers in Fig.~\ref{fig:negsel}(a,b).  Note that for $s>0$, the probabilities converge rapidly to their steady-state value.  Conversely, for $s<0$, the data points at different $t$ do not overlap in Fig.~\ref{fig:negsel}(a,b), indicating that the probabilities have not yet converged.  We check that the survival probability has not yet approached a steady-state for small negative $s$ in Fig.~\ref{fig:negsel}(c,d) for both circular and spherical frontiers.

\section*{References}

\bibliography{SurvivalELSBib}

\end{document}